\documentclass[aps,preprint,pra]{revtex4-1}
\usepackage{graphicx}
\usepackage{amssymb}
\usepackage{amsmath}
\usepackage{color}
\usepackage{enumerate}
\usepackage{bm}
\usepackage[utf8]{inputenc}
\begin{document}
%\iffalse
%\label{key}\title{ Disorder-Induced Complex Magnetization Dynamics in Planar Nanostructure}
\title{Magnetization Reversal in Two-dimensional Ensemble of Nanoparticles with Positional Defects}
\author{Manish Anand}
\email{itsanand121@gmail.com}
\affiliation{Department of Physics, Bihar National College, Patna University, Patna-800004, India.}

\date{\today}
\begin{abstract}
We study relaxation behaviour in the two-dimensional assembly of magnetic nanoparticles (MNPs) with aligned anisotropy axes and positional defects. The anisotropy axes orientation and disorder strength is changed by varying $\alpha$ and $\Delta$, respectively. The magnetization decay does not depend on the aspect ratio $A^{}_r$ of the system and $\Delta$ for small dipolar interaction strength $h^{}_d=0.2$. Remarkably, the magnetization decays rapidly for considerable $h^{}_d$ with negligible $\Delta$ and $A_r=1.0$. The dipolar interaction of enough strength promotes antiferromagnetic coupling in square ensembles of MNPs. There is a prolonged magnetization decay for large $\Delta$ because of enhancement in ferromagnetic coupling. Notably, magnetization relaxes slowly for $\alpha<\alpha^\star$ even with moderate $h^{}_d$ and a significant $A^{}_r$. Interestingly, the slowing down of the magnetic relaxation shifts to a lower $\alpha^{\star}$ with $h^{}_d=1.0$. The magnetization ceases to relax for $\alpha\leq60^\circ$ and $h^{}_d\leq0.6$ due to large shape anisotropy with $A^{}_r=400.0$. Remarkably, a majority of the magnetic moment reverses its direction by $180^\circ$ for $\alpha>60^\circ$ and large $h^{}_d$, resulting in the negative magnetization. The effective N\'eel relaxation time $\tau^{}_N$ also depends strongly on these parameters. $\tau^{}_N$ depends weakly on $\alpha$ and $\Delta$ for $h^{}_d\leq0.2$, irrespective of $A^{}_r$. On the other hand, $\tau^{}_N$ decreases with $\alpha$ for significant $h^{}_d$ provided $\alpha$ is greater than $45^\circ$ because of antiferromagnetic coupling dominance. In a highly anisotropic system, there is an enhancement in $\tau^{}_N$ with $\alpha$ ($\leq30^\circ$) even with moderate $h^{}_d$. While for $\alpha>30^\circ$, $\tau^{}_N$ decreases with $\alpha$. These observations are useful in novel materials, spintronics based applications, etc. 
\end{abstract}
\maketitle
%\fi
\section{Introduction}
Two-dimensional assemblies of magnetic nanoparticles (MNPs) have received significant attention due to their exciting properties and numerous technological applications~\cite{leo2018,srivastava2014,gallina2020,puntes2001,moya2015,wang2019,sanchez2020,varon2013dipolar,peng2018,manish2021}. These applications include magnetic hyperthermia mediated cancer therapy, targeted drug delivery, magnetic resonance imaging, data storage, etc. The magnetic properties of these systems are complex and extremely sensitive to various parameters of interest such as particle shape and size, anisotropy strength, spatial distribution, dipolar interaction, etc.~\cite{farhan2013,liu2002,puntes2004}. The understanding of magnetic relaxation characteristics are of paramount importance for their efficient usages in such applications~\cite{fabris2019,perez2002,pisane2015}. The N\'eel relaxation time is one of the essential quantifiers of magnetic relaxation. The precise knowledge of relaxation time is also of great significance in various applications such as data storage, magnetic hyperthermia, etc.       

The N\'eel-Brown theory of superparamagnetism correctly explains the relaxation properties in the case of single nanoparticle or extremely dilute assembly~\cite{hergt2009,wernsdorfer1997}. However, the dipolar interaction strongly affects the magnetization dynamics behaviour in sufficiently dense assembly~\cite{figueiredo2007,dejardin2011}. The dipolar interaction is long ranged and anisotropic in nature. As a consequence, it can induce ferromagnetic or antiferromagnetic coupling depending on the relative position of the nanoparticle in an assembly. Therefore, the relaxation characteristics are severely affected because of the dipolar interaction~\cite{rizzo1999,denisov2002,hiroi2011,kesserwan2011,kuncser2019,shtrikman1981,morup2010,dormann1988}. For example, Denisov {\it et al.} investigated the dipolar interaction dependence of magnetic relaxation in two-dimensional ensembles of MNPs using mean field approximations~\cite{denisov2002}. They observed two distinct relaxation time because of dipolar interaction. Hiroi {\it et al.} studied the magnetization dynamics in an assembly of hematite (Fe$^{}_2$O$^{}_3$) as a function of interparticle seperation or dipolar interaction strength~\cite{hiroi2011}. The system exhibits super spin glass-like properties in the case of sufficiently large interaction strength. Kesserwan {\it et al.} investigated the effect of dipolar interaction on the magnetization dynamics in MNPs ensembles using Fokker Planck formalism~\cite{kesserwan2011}. The dipolar interaction is found to induce fastening in magnetization relaxation. Kuncser {\it et al.} analyzed the magnetic relaxation as a function of dipolar interaction~\cite{kuncser2019}. The energy barrier gets enhanced due to the dipolar interaction and hence the relaxation time. Shtrikmann and Wohl­farth investigated the dipolar interaction dependence of the relaxation time~\cite{shtrikman1981}. They observed an enhancement in relaxation time with an increase in interaction strength. M{\o}rup {\it et al.} and Dormann {\it et al.} also found an elevation in relaxation time because of the dipolar interaction~\cite{morup2010,dormann1988}.

The influence of dipolar interaction on relaxation characteristics
also depends strongly on the spatial distribution of nanoparticles. For instance, the energy barrier gets enhanced and hence the relaxation time due to the dipolar interaction~\cite{anand2021thermal,salvador2021}. On the other hand, the dipolar interaction increases the relaxation rate in the square assembly of MNPs~\cite{anand2021magnetic}. Patrick Ilg studied the magnetization relaxation of multicore nanoparticles dispersed in viscous solvent using computer simulations~\cite{ilg2017}. In the case of dense random cluster of MNPs, the magnetic relaxation curve follows exponentially decaying function for moderate interaction strength. L\'opez {\it et al.} analyzed the relaxation mechanism in polycrystalline sample of
La$^{}_{0.5}$Ca$^{}_{0.5}$MnO$^{}_3$~\cite{lopez2001}. The magnetization decay curve is found to follow the logarithmic function. Osaci {\it et al.} investigated the relaxation properties as a function of dipolar interaction strength using numerical simulations~\cite{osaci2006}. The relaxation time gets enhanced with an increase in interaction strength.
%In the case of randomly  placed particles, the dipolar interaction induces spin  glass like properties. 
The anisotropy axes orientations also play an important role in determining the magetic behaviour in MNPs ensembles~\cite{eloi2014,thorkelsson2015,anand2020hysteresis,yoshida2017effect,wang2008fept,khurshid2021tuning}. Sharma {\it et al.} investigated the anisotropy axes orientation effect on the hysteresis properties in ordered MNPs assembly~\cite{sharma2010}. The coercive field and remanence magnetization get enhanced in the highly anisotropic system. Usov {\it et al} also observed larger coercivity for alligned anisotropy axes~\cite{usov2013}. Conde-Lebor\'an {\it et al.} probed the hysteresis characteristis of interacting nanoparticles as a function of the degree of collinearity of anisotropy axes~\cite{conde2015}. They found the maximum heat dissipation in the case of aligned anisotropy axes.
In a recent work, we have shown the relaxation time can be modulated by just varying the angle between the anisotropy axis and chain axis of MNPs~\cite{anand2019}. In a another study, we have shown the N\'eel relaxation time can be enhanced or decreased by varying the anisotropy axes in the prefectly ordered system~\cite{anand2021relaxation}.   

%Because of the long-range and anisotropic nature, the effect of dipolar interaction on the relaxation also depends strongly on the geometrical arrangement and anisotropy axes orientations [24–32].
The above discussion clearly indicates the dipolar interaction and anisotropy axes orientation play a crucial role in determining various magnetic properties of interest in MNPs ensembles. Therefore, the precise control over dipolar interaction strength and orientation of anisotropy axes could be of great importance in tuning diverse systematic properties. They find significant usages in various applications such as plasmonics, data storage, sensors, photovoltaics, etc.~\cite{aldaye2008,boal2000,huynh2002}. Therefore, the fabrication of such system with high degree of structural and orientational order have gained momentum in recent years~\cite{fava2008,ryan2006,ploshnik2010}. However, such experimentally fabricated MNPs ensembles are found to have various types of defects, which further complicates the magnetic response. Of all possible defects, the out-of-plane positional disorder is often observed in these fabricated samples. In such a situation, all the constituent MNPs do not lie in a single plane, some of them also scatter normal to the plane of sample. Consequently, the complex and unexpected systematic properties are observed due to interplay among the anisotropy axes orientation, dipolar interaction, and out-of-plane disorder. Therefore, it is essential to understand the magnetic properties of such system and provide a theoretical framework to explain such unexpected behaviour. These studies could be extremely useful in controlling and modulating the systematic properties of such system to harness
their unique anisotropic properties of technological importance. Thus motivated, we investigate the magnetic relaxation in two-dimensional assembly of MNPs as a function of system size and aspect ratio, dipolar interaction, anisotropy axes orientations and out-of-plane disorder strength using kinetic Monte Carlo simulations. 

The rest of the article is organized as follows. In Section II, we present and discuss our theoretical framework and relevant energy terms. We also briefly discuss the procedures of numerical 
simulations. In Section III, we describe the numerical results. Finally, in Section IV, we provide the summary of the present work.
%summarize the main results presented in this work.
%{\color{red} Understanding and controlling the properties of the magnetic nanostructures is paramount for an efficient application.}

\section{Theoretical Framework}
\begin{figure}[!htb]
	\centering\includegraphics[scale=0.450]{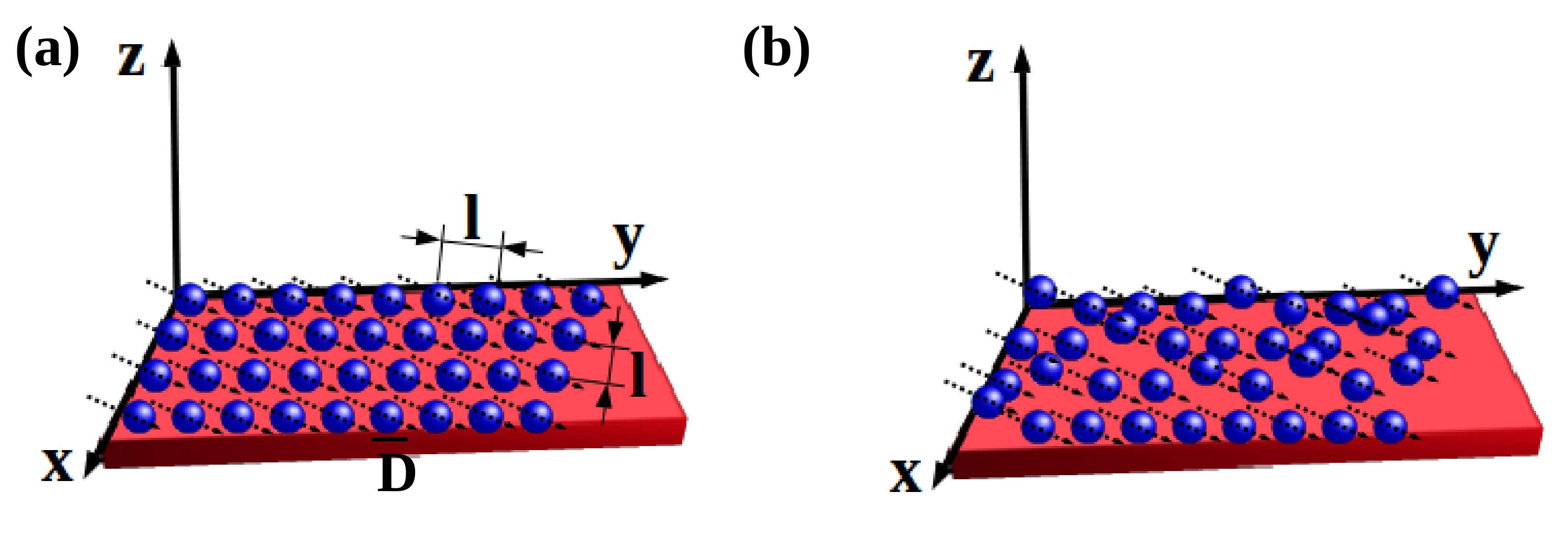}
	\caption{(a) Schematic of the two-dimensional ensemble of magnetic nanoparticles assembled in the $xy$-plane. $l$ is the lattice constant, and $D$ is the particle diameter. The anisotropy axes are denoted by dashed lined arrows, making an angle $\alpha$ with the $y$-axis. (b) Schematic of the nanoparticles assembly with the out-of-plane disorder. A few nanoparticles are dispersed normal to the xy-plane.}
	\label{figure1}
\end{figure}

We consider a two-dimensional ($l^{}_x\times l^{}_y$) ensemble of spherical and monodisperse nanoparticles in the $xy$-plane with some of them dispersed normal to the sample plane (along $z$-axis) as depicted in Fig.~(\ref{figure1}). Let the particle diameter be $D$ and the lattice spacing be $l$. Let the magnetocrystalline anisotropy constant be $K^{}_{\mathrm {eff}}$ and particle volume $V=\pi D^{3}/6$. So, the anisotropy energy of single particle can be expressed as~\cite{anand2016spin,carrey2011}.

%The nanoparticles are assembled in the two-dimensional lattice ($l^{}_x\times l^{}_y$) in the $xy$-plane, with a few of them  scattered along the $z$-axis, as shown in the schematic Fig.~(\ref{figure1}). 
%The particle has a diameter $D$, and the lattice spacing is $l$, as depicted in Fig.~(\ref{figure1}).
%We consider an assembly of spherically shaped and monodisperse magnetic nanoparticle arranged in a two-dimensional lattice in the $xy$-plane with system dimension $L^{}_x\times L^{}_y$. 
%The particle has a diameter $D$, and the lattice spacing is $a$, as shown in Fig.~\ref{figure1}(a). 
%The  magnetocrystalline anisotropy constant associated with each nanoparticle is $K^{}_{\mathrm {eff}}$; $V=\pi D^{3}/6$ is the particle volume. 

%Therefore, the magnetocrystalline energy associated with such a nanoparticle is given by~\cite{anand2016spin,bedanta2008supermagnetism}
\begin{equation}
	E=K^{}_\mathrm {eff} V\sin^2\Psi.
	\label{barrier}
\end{equation}
Here $\Psi$ is the angle between the anisotropy or the easy axis and the magnetic moment $\mu=M^{}_sV$; $M^{}_s$ is the saturation magnetization. The above energy function is a symmetric double well with two minima $E^{o}_1$ (say) and $E^{o}_2$ (say) at $\Psi=0$ and  $\Psi=\pi$, respectively. These are seperated by an energy maximum $E^{o}_3 (\mathrm {say})= K^{}_\mathrm {eff} V$ at $\Psi=\pi/2$. The latter is also termed as the energy barrier. In the presence of sufficient temperature $T$, each magnetic moment associated with a nanoparticle fluctuates between these two minima with oppositely directed magnetization by surmounting the energy barrier $K^{}_\mathrm {eff} V$. Such magnetization fluctuation is characterized by a time scale known as the N\'eel relaxation time $\tau^{o}_N$, which is given by the Arrhenius relation~\cite{anand2019,carrey2011}
%The magnetic moment changes its position from $\Phi=0$ to $\pi$ and vice versa by surmounting the energy barrier in the presence of sufficient thermal fluctuations. The mean time taken by the magnetic moment to perform such flips is termed as the N\'eel relaxation time $\tau^{0}_N$~\cite{anand2019,carrey2011}
\begin{equation}
	\tau^{o}_N=\tau^{}_o\exp(K^{}_\mathrm {eff}V/k^{}_BT).
	\label{Neel}
\end{equation}
$\tau^{}_o$ is the the microscopic limiting relaxation time usually related to the attempt frequency $\nu^{}_o\approx10^{10}$ $s^{-1}$ as  $\tau^{}_o=(2\nu_o)^{-1}$. $k^{}_B$ is the Boltzmann constant. Eq.~(\ref{Neel}) correctly estimates the relaxation time of single particle or weakly interacting nanoparticles ensembles.
%The above relation is applicable to an isolated particle or the extremely dilute assembly of MNPs.
%Eq.~(\ref{Neel}) is valid for a single nanoparticle.

%applicable for non-interacting MNPs.
In an assembly of MNPs, the single particle energy barrier gets drastically modified because of the dipolar interaction. The corresponding interaction energy $E^{}_{\mathrm {dipolar}}$ can be evaluated using the following expression~\cite{ANAND2021168461,bupathy2019}
%The single-particle energy barrier is primarily modified due to the long-ranged dipolar interaction. The dipolar interaction energy $E^{}_{\mathrm {dip}}$ can be calculated by the following relation~\cite{kechrakos1998,anand2020hysteresis}
%\begin{equation}
%\label{dipole}
%E_{dd}(s)= -\frac{1}{4\pi \mu_{o}}\left(\frac{3\left(\vec{\mu_{i}}\cdot\vec{s}\right)\left(\vec{\mu_{j}}\cdot \vec{s}\right)}{s^{5}} -\frac{\vec{\mu_{i}}\cdot \vec{\mu_{j}} }{s^3}\right),
%\end{equation}
\begin{equation}
	\label{dipole}
	E^{}_{\mathrm {dipolar}}=\frac{\mu^{}_o \mu^{2}}{4\pi l
		^3}\sum_{j,\ j\neq i}\left[ \frac{\hat{\mu_{i}}\cdot\hat{\mu_{j}}-3\left(\hat{\mu_{i}}\cdot\hat{r}_{ij}\right)\left(\hat{\mu_{j}}\cdot\hat{r}_{ij}\right)}{(r^{}_{ij}/l)^3}\right].
\end{equation}
$\mu_{o}$ is the permeability of free space. $\hat{\mu_{i}}$ and $\hat{\mu_{j}}$ are the unit vectors for the magnetic moments at $i^{th}$ and $j^{th}$ position, respectively.
%the nanoparticle at $i^{th}$ and $j^{th}$ position has magnetic moment  $\vec{\mu}_{i}$ and $\vec{\mu}_{j}$, respectively. 
$r^{}_{ij}$ is the distance between $\hat{\mu_{i}}$ and $\hat{\mu_{j}}$.
%The particle has a magnetic moment $\mu=M^{}_sV$, $M^{}_s$ is the saturation magnetization. 
%$\mu^{}_o\vec{H}^{}_{\mathrm {dip}}$ is the dipolar field due to all other nanoparticles, present in the system. 
The corresponding dipolar field  $\mu^{}_o\vec{H}^{}_{\mathrm {dipolar}}$ can be expressed as~\cite{tan2014,azeggagh2007}
\begin{equation}
	\mu^{}_{o}\vec{H}^{}_{\mathrm {dipolar}}=\frac{\mu_{o}\mu}{4\pi l^3}\sum_{j,j\neq i}\frac{3(\hat{\mu}^{}_j \cdot \hat{r}_{ij})\hat{r}^{}_{ij}-\hat{\mu^{}_j} }{(r_{ij}/l)^3}.
	\label{dipolar1}
\end{equation}
%Here $\mu_{j}$ is the magnetic moment vector of the $j^{th}$ particle, $s$ is the center-to-center separation between $\mu_{i}$ and $\mu_{j}$; and the $\mu_{o}$ is the permeability of free space.
It is evident from Eq.~(\ref{dipole}) and Eq.~(\ref{dipolar1}) that the long-ranged dipolar interaction varies as $1/r^{3}_{ij}$. Therefore, we can model the variation of dipolar interaction strength by introducing a control parameter $h^{}_d=D^{3}/l^{3}$~\cite{tan2010}. The physics of non-interacting particles can be modelled with $h^{}_d=0.0$. On the other hand, $h^{}_d=1.0$ represents the strongest interacting system. %As $h^{}_d=0$ implies $l\to\infty$, it mimics the non-interacting situation. On the other hand, the strongest dipolar interacting case can be modelled by $h^{}_d=1.0$.
The total energy of the system can be caculated using the following expression~\cite{tan2014,anand2019}  
%We can now write the total energy $E$ of the underlying system as~\cite{tan2014,anand2019}
\begin{equation}
	E=K^{}_{\mathrm {eff}}V\sum_{i}\sin^2 \Psi^{}_i+\frac{\mu^{}_o\mu^2}{4\pi l^3}\sum_{j,\ j\neq i}\left[ \frac{\hat{\mu_{i}}\cdot\hat{\mu_{j}}-{3\left(\hat{\mu_{i}}\cdot\hat{r}_{ij}\right)\left(\hat{\mu_{j}}\cdot\hat{r}_{ij}\right)}}{(r_{ij}/l)^3}\right]
\end{equation}
%Here $\theta^{}_i$ is the angle between the anisotropy axis and the $i^{th}$ magnetic moment of the system. 

The dipolar interaction significantly modifies the single-particle energy function (Eq.~(\ref{barrier})). Consequently, the functional form of the energy barrier becomes asymmetric. The energy extrema also gets modified because of dipolar interaction. Let the new energy minima be $E^{}_1$ and $E^{}_2$. Let us also assume the modified energy maxima to be $E^{}_3$.
%Consequently, the energy extrema of the nanoparticle also get altered. Let $E^{}_1$ and $E^{}_2$ be new energy minima and $E^{}_3$ be the modified energy maximum. %Consequently, the magnetic moment chages its position from $E^{}_1$ to $E^{}_2$ and vice versa. 
In such a case, the rate $\nu^{}_1$ with which the magnetic moment changes its orientation from $E^{}_1$ to $E^{}_2$ by surmounting  $E^{}_3$ can be computed using the following formula~\cite{hanggi1990}

%As the energy barrier seen by a magnetic moment gets alterted by the dipolar interaction, the 
%The dipolar interaction modifies the single-particle energy barrier given by Eq.~(\ref{barrier}) is modified due to the dipolar interaction. Consequently, the single-particle energy function defined  becomes asymmetric, as depicted in Fig.~\ref{figure1}(c). The modified energy function
%The modified energy function has single energy minimum when the dipolar field is larger than the anisotropy field $H^{}_K=2K^{}_{\mathrm {eff}}/M^{}_s$~\cite{carrey2011}.
%two new energy minima $E^{}_1$ and $E^{}_2$, and a maxima $E^{}_3$ for  $|\mu^{}_{o}\vec{H}^{}_{\mathrm {dip}}|<H^{}_K$, as shown in Fig.~\ref{figure1}(c). Therefore, the rate $\nu^{}_1$ at which the magnetic moment goes from $E^{}_1$ to $E^{}_2$ via $E^{}_3$ is expressed as~\cite{hanggi1990}
\begin{equation}
	\nu^{}_1=\nu^{0}_{1}\exp\bigg(-\frac{E^{}_3-E^{}_1}{k^{}_BT}\bigg)
\end{equation}
Similarly, the following relation gives the jump rate $\nu^{}_2$ to switch the direction of the magnetic moment from $E^{}_2$ to $E^{}_1$~\cite{hanggi1990}
\begin{equation}
	\nu^{}_2=\nu^{0}_2\exp\bigg(-\frac{E^{}_3-E^{}_2}{k^{}_BT}\bigg),
\end{equation} 
where $\nu^{0}_{1}=\nu^{0}_{2}=\nu^{}_{o}$. 

We investigate the magnetic relaxation in two-dimensional assembly ($l^{}_x\times l^{}_y$) of MNPs as a function of anisotropy axis orientation, out-of-plane disorder, dipolar interaction strength $h^{}_d$ and aspect ratio $A^{}_r=l^{}_y/l^{}_x$ using numerical simulations. The anisotropy axis of each constituent nanoparticle in the assembly is assumed to make an angle $\alpha$ with the $y$-axis as shown in the schematic Fig.~(\ref{figure1}). We define a control paramater $\Delta$ to model the effect of the out-of-plane disorder strength variation on the magnetization dynamics. It is defined as $\Delta(\%)=n^{}_z/N$, $N$ is the total number of MNPs in the assembly and $n^{}_z$ is the number of particles dispersed normal to the sample plane (along $z$-axis). The kinetic Monte Carlo (kMC) algorithm is implemented for computer simulations. It is describied in the greater detail in the work Tan {\it et al.} and Anand {\it et al.}~\cite{tan2014,anand2019}. Therefore, we do not reiterate it here to avoid duplication. We have also described the implemented kMC algoirthm in our recent works~\cite{anand2021tailoring,ANAND2021168461}. We now describe the protocol used for investigation of the magnetic relaxation. We apply extremely large constant magnetic field of strength of $30$ T along the $y$-axis to saturate the magnetization. We then divide the total simulation time into 2000 equal step. The external biasing magnetic is switched off at time $t=0$. Therefater, we study the magnetization dynamics using the kMC algorithm. The simulated magnetization decay $M(t)/M^{}_s$ vs $t$ curve is then fitted with $M(t) = M^{}_s \exp(-t/\tau^{}_N)$ to extract the effective N\'eel relaxation time $\tau^{}_N$ of the underlying system.

%To evaluate the time dependence of magnetization, we apply a huge constant magnetic field $\mu_oH_{\mathrm{max}}$ of strength 20 T along $y$-direction to align all the magnetic moments along the external field direction. After that, the total simulation time is divided into 2000 equal steps, and the applied magnetic field $\mu_oH_{\mathrm{max}}$  is switched off at time $t=0$. We then evaluate the time evolution of magnetization using the kMC algorithm. We fit the so-obtained magnetization decay $M(t)/M^{}_s$ vs $t$ curve with $M(t) = M^{}_s \exp(-t/\tau^{}_N)$ to extract the effective N\'eel relaxation time $\tau^{}_N$.

\section{Simulations Results}
We have considered the following values of system parameters for numerical simulations: $D=8$ nm, $K^{}_{\mathrm {eff}}=13\times10^{3}$ $\mathrm {Jm^{-3}}$, $M^{}_s=4.77\times10^{5}$ $\mathrm {Am^{-1}}$. The total number of nanoparticles is taken as $N=400$.  We have taken five typical values of system sizes $l^{}_x\times l^{}_y=20\times20$, $10\times40$, $4\times100$, $2\times200$, and  $1\times400$. These correspond to the aspect ratio $A^{}_r(=l^{}_y/l^{}_x)$: $1.0$, 4.0, 25.0, 100.0, 400.0, respectively. We have varied the out-of-plane disorder strength $\Delta$ from 0 to 50\%. The dipolar interaction strength $h^{}_d$ is changed between 0.0 to 1.0. We have also varied the orientation of anisotropy axis $\alpha$ from 0 to $90^\circ$ and temperature $T$ is taken as 300 K.
%$\Delta (\%)$ is varied between 5 to 50 to investigate the dependence of the out-of-plane disorder on  magnetic relaxation extensively. We have also varied the dipolar interaction strength $h^{}_d$ from 0 to 1.0. The anisotropy axes are taken as randomly oriented in the three-dimensional space.

\begin{figure}[!htb]
	\centering\includegraphics[scale=0.35]{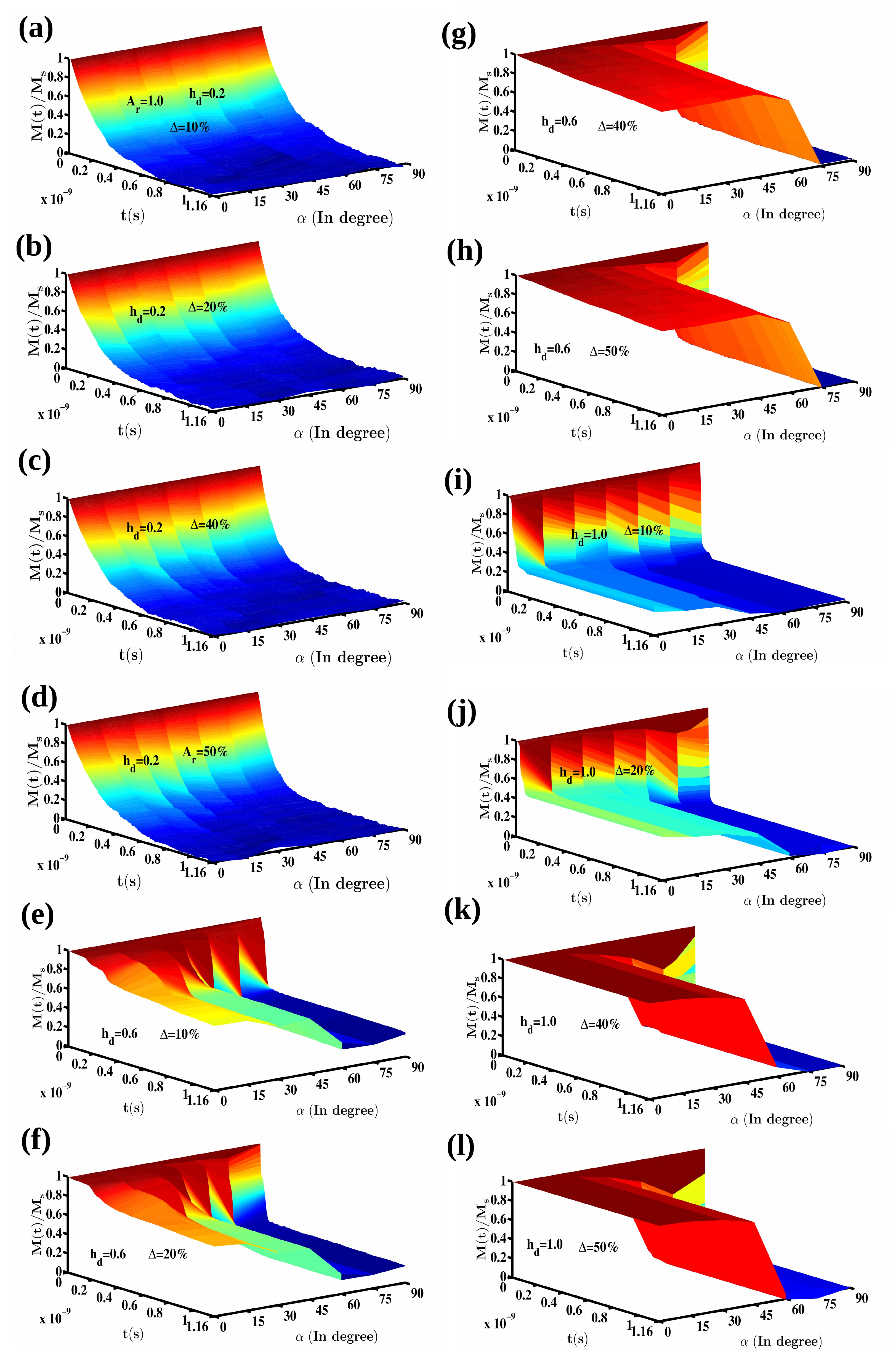}
	\caption{Magnetization decay curve as a function of time $t$ and anisotropy axes orientation angle $\alpha$ with $A^{}_r=1.0$. We have considered three representative values of $h^{}_d=0.2$ [(a)-(d)], 0.6 [(e)-(h)], and 1.0 [(i)-(l)]. Four typical values of $\Delta=10\%$, 20\%, 40\%, and 50\% are also considered. The magetic relaxation is perfectly exponentially decaying for $h^{}_d=0.2$, irrespective of $\alpha$ and $\Delta$. The magnetization decays rapidly for large $h^{}_d$ and negligible $\Delta$.}
	\label{figure2}
\end{figure}

We first study the magnetic relaxation behaviour in square arrays of nanoparticles as a function of dipolar interaction strength, out-of-plane disorder strength and anisotropy axis orientation. In Fig.~(\ref{figure2}), we plot the magnetization decay $M(t)/M^{}_s$ vs $t$ curve as a function of $h^{}_d$, $\alpha$ and $\Delta$ with $A^{}_r=1.0$. We have considered three representative values of $h^{}_d=0.2$, 0.6, and 1.0. We have also taken four typical values of $\Delta=10\%$, 20\%, 40\%, and 50\%. The anisotropy axes orientational angle $\alpha$ is varied from 0 to $90^\circ$. There is a smooth decay of magnetization for weakly interacting MNPs ($h^{}_d=0.2$). In such a case, the functional form of the time evolution of the magnetization-decay is exponentially decaying and  depends weakly on the $\alpha$ and $\Delta$. On the other hand, the magnetic relaxation characteristics depend strongly on the anisotropy axes orientation and out-of-plane disorder strength for strongly dipolar interacting MNPs. Remarkably, the magnetization decays rapidly with large dipolar interaction strength and negligible value of disorder. Such rapid decay of magnetization does not depend on $\alpha$. It is because the dipolar interaction of sufficient strength induces antiferromagnetic coupling in square ensembles of MNPs, resulting in faster magnetization decay. These observations are in perfect qualitative agreement with the work of De'Bell {\it et al.}~\cite{de1997}. In contrast, there is a slowing down of the magnetization relaxation for considerable disorder strength. Notably, the magnetization ceases to relax for $\alpha\leq\alpha^{\star}$. On the other hand, there is a fastening in the magnetization relaxation for $\alpha>\alpha^{\star}$. Surprisingly, there is an increase in $\alpha^{\star}$ with $\Delta$. It can be explained by the fact that the out-of-plane disorder of significant strength instigates magnetic moments to interact ferromagnetically even in the square assembly of MNPs. Consequently, the magnetization ceases to relax in such cases.

\begin{figure}[!htb]
\centering\includegraphics[scale=0.35]{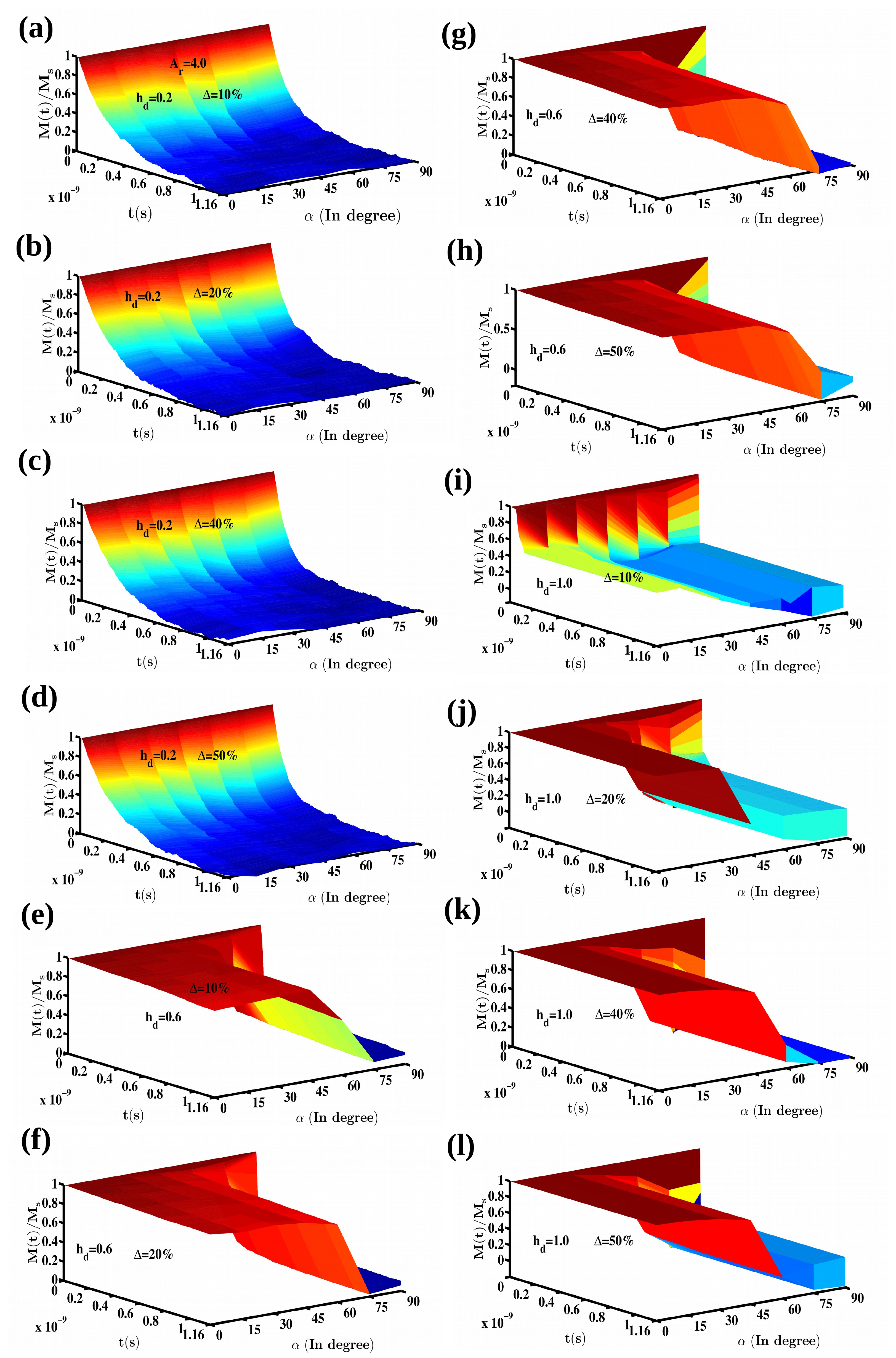}
\caption{Magnetic relaxation as a function of time $t$ and anisotropy axes orientation angle $\alpha$ with $A^{}_r=4.0$. Three representative values of $h^{}_d=0.2$ [(a)-(d)], 0.6 [(e)-(h)], and 1.0 [(i)-(l)] are considered. Four typical values of $\Delta=10\%$, 20\%, 40\%, and 50\% are also considered. The functional form of the magnetization decay depends weakly on $\alpha$ and $\Delta$ for $h^{}_d=0.2$. The magnetization relaxes slowly for $\alpha<75^\circ$ and moderate $h^{}_d$.}
\label{figure3}
\end{figure}

Next, we investigate the magnetization-decay characteristics in the rectangular arrays of MNPs. Fig.~(\ref{figure3}) shows the time evolution of the magnetization decay $M(t)/M^{}_s$ vs $t$ curve with $A^{}_r=4.0$. We have used the same set of values for $\alpha$, $h^{}_d$, and $\Delta$ as in Fig.~(\ref{figure2}). In the case of small dipolar interaction strength ($h^{}_d=0.2$), the magnetization decays smoothly even with the rectangular assembly of MNPs. In this case, the functional form of the magnetization relaxation also depends very weakly on $\alpha$ and $\Delta$. Interestingly, the magnetization ceases to relax for $\alpha\leq75^\circ$ even with a moderate value of interaction strength $h^{}_d=0.6$ and a small value of $\Delta=10\%$. On the other hand, there is a rapid magnetization decay for $\alpha>75^\circ$, irrespective of $\Delta$. It is because the antiferromagnetic coupling is the very large in such a case. Remarkably, the slowing down of the magnetic relaxation shifts to a lower value of $\alpha=\alpha^{\star}$ with largest dipolar interaction strength $h^{}_d=1.0$. The out-of-plane disorder enhances the value of $\alpha^\star$ by increasing the ferromagnetic coupling among the magnetic moments. 

\begin{figure}[!htb]
\centering\includegraphics[scale=0.35]{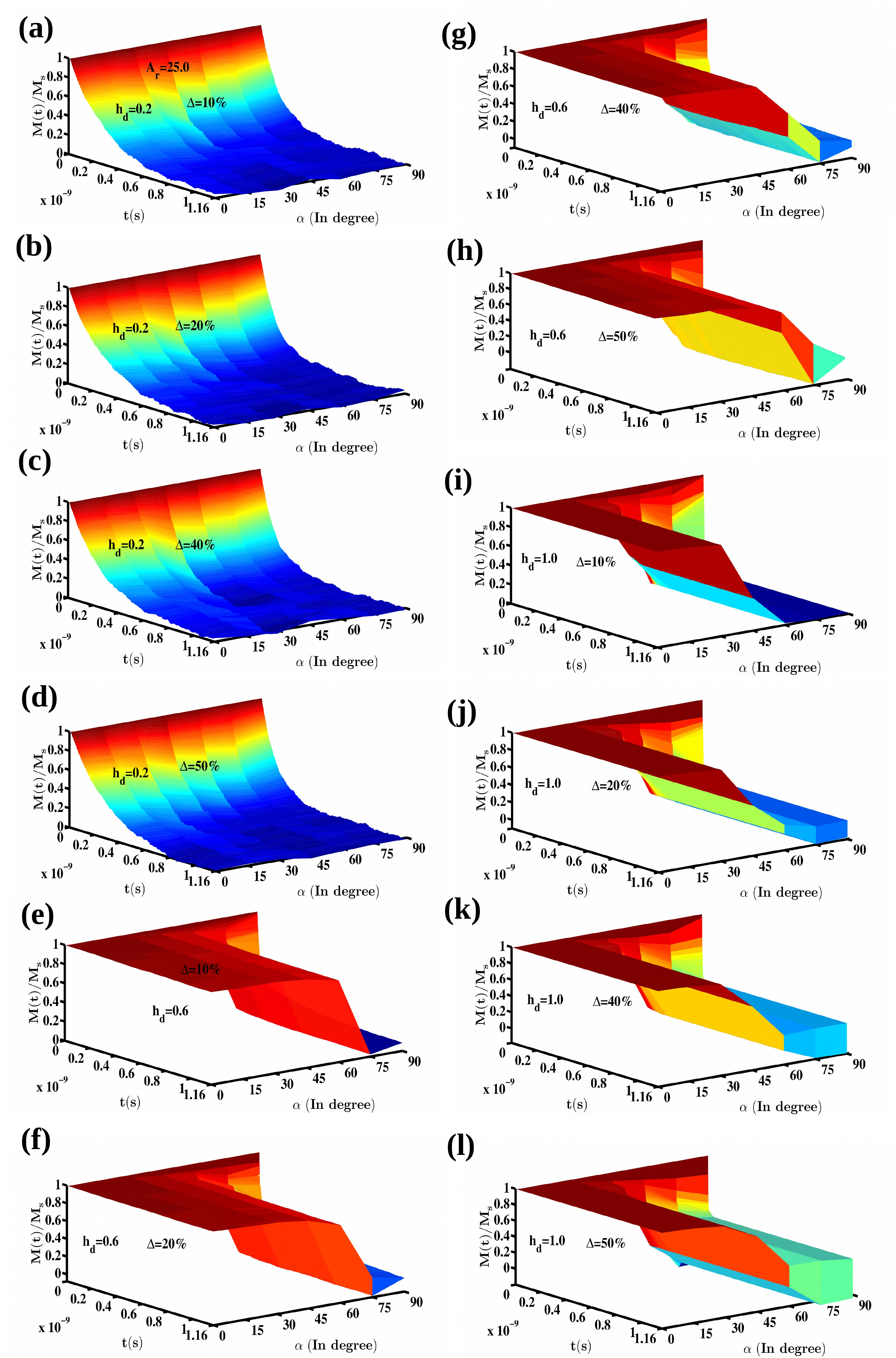}
\caption{Magnetization relaxation as a function of time $t$ and anisotropy axes orientation angle $\alpha$ with $A^{}_r=25.0$. We consider three representative values of $h^{}_d=0.2$ [(a)-(d)], 0.6 [(e)-(h)], and 1.0 [(i)-(l)]. Four typical values of $\Delta=10\%$, 20\%, 40\% are also taken. The magnetization decay depends weakly on $\alpha$ and $\Delta$ for small $h^{}_d=0.2$. The magnetic relaxation slows down for $\alpha\leq75^\circ$ for $h^{}_d=0.6$, even with small $\Delta$.}
\label{figure4}
\end{figure}

\begin{figure}[!htb]
\centering\includegraphics[scale=0.35]{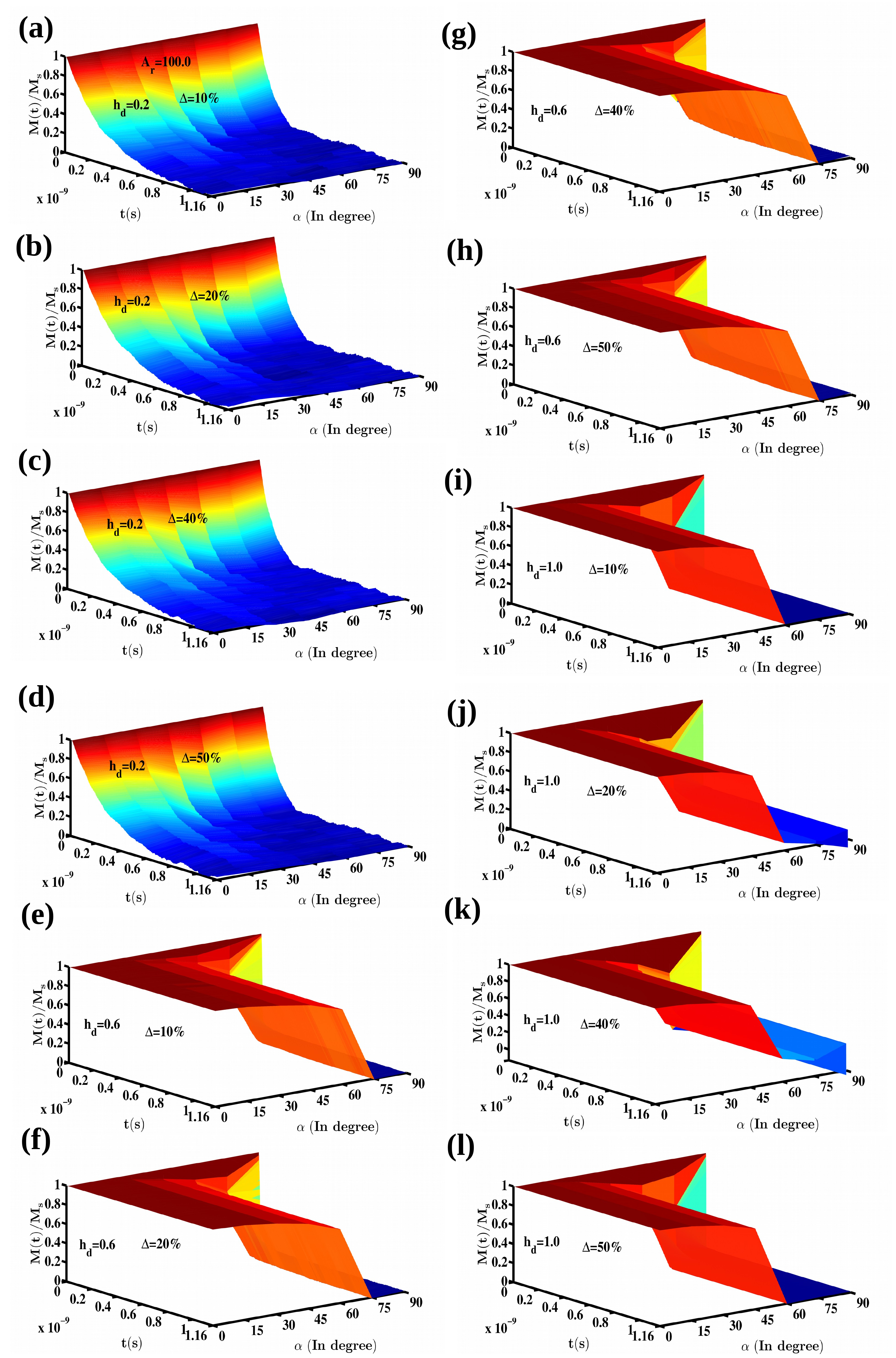}
\caption{Magnetization decay $M(t)/M^{}_s $ vs $t$ and anisotropy axes orientation angle $\alpha$ with $A^{}_r=100.0$. We have considered three representative values of $h^{}_d=0.2$ [(a)-(d)], 0.6 [(e)-(h)], and 1.0 [(i)-(l)]. Four typical values of $\Delta=10\%$, 20\%, 40\%, and 50\% are also considered. The magnetic relaxation depends weakly on $\alpha$ and $\Delta$ for small $h^{}_d$. While for significant $h^{}_d$ and $\Delta$, the magnetization relaxation depends strongly on $\alpha$.}
\label{figure5}
\end{figure}

We then analyze the magnetization decay behaviour in a system with a huge aspect ratio. We plot the time evolution of the magnetization-decay $M(t)/M^{}_s$ vs $t$ curve as a function of $\alpha$ for $A^{}_r=25.0$ and 100.0 in Fig.~(\ref{figure4}) and Fig.~(\ref{figure5}), respectively. Even with a highly anisotropic system, the magnetization decays smoothly for weakly interacting MNPs, irrespective of $\alpha$ and $\Delta$. For moderate dipolar interaction strength ($h^{}_d=0.6$), the magnetic relaxation slows down for $\alpha<75^\circ$ even with small out-of-plane disorder strength $\Delta$. It is because the ferromagnetic coupling is enhanced because of out-of-plane disorder strength. In the case of the largest dipolar interaction strength ($h^{}_d=1.0$), the slowing down of magnetization-decay occurs relatively for smaller anisotropy axis orientation $\alpha$ with $A^{}_r=25.0$  compared to $A^{}_r=100.0$. It can be explained by the fact that the shape anisotropy becomes prominent in the highly anisotropic system, which strengthens the ferromagnetic coupling. Consequently, the magnetic moments tend to align along the long axis of the system, resulting in prolonged magnetization decay even with a large value of $\alpha$. 

\begin{figure}[!htb]
\centering\includegraphics[scale=0.35]{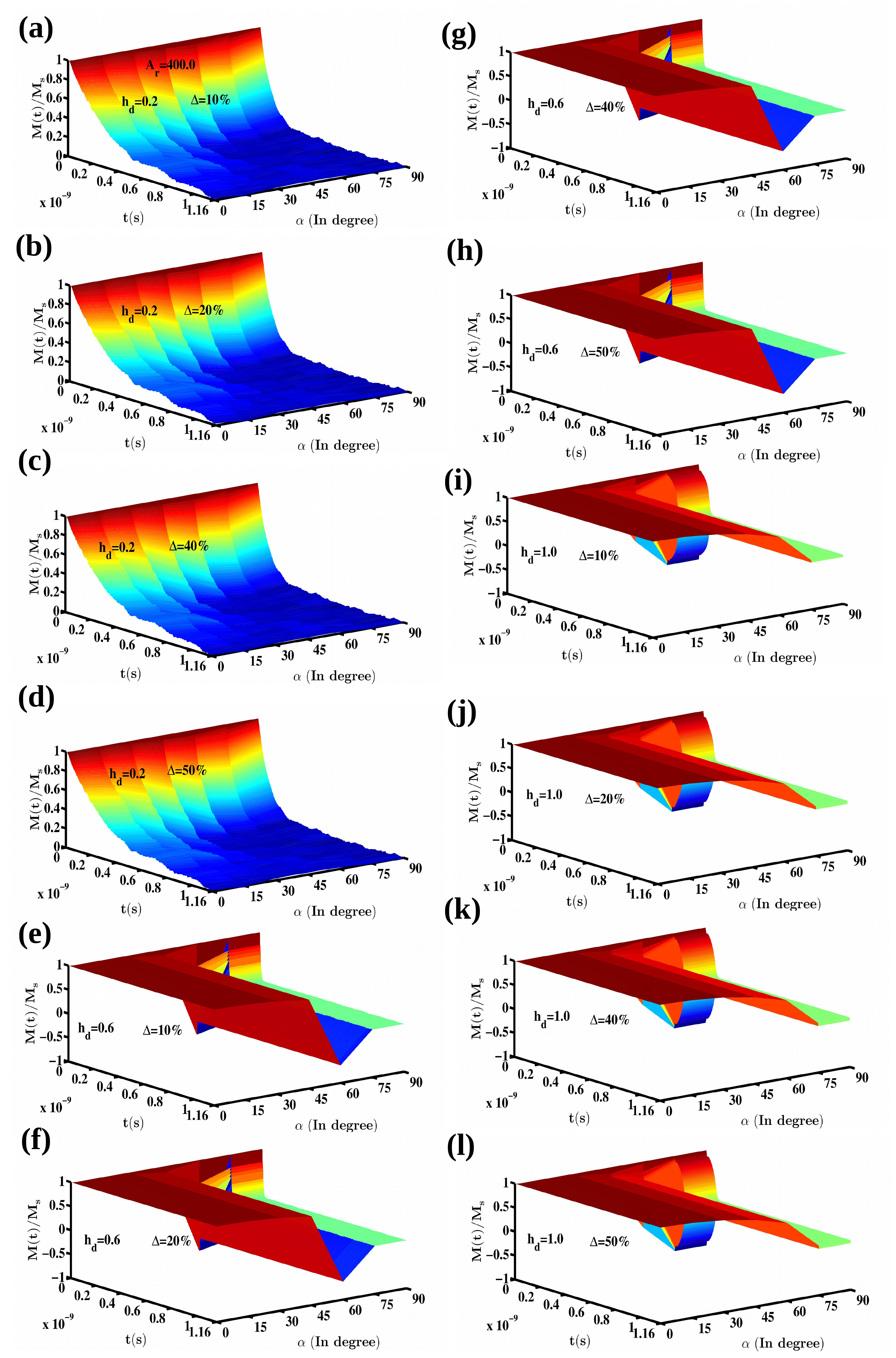}
\caption{The magnetization relaxation $M(t)/M^{}_s $ vs $t$ and anisotropy axes orientation angle $\alpha$ with $A^{}_r=400.0$. We have taken three typical values of $h^{}_d=0.2$ [(a)-(d)], 0.6 [(e)-(h)], and 1.0 [(i)-(l)]. Four representative values of $\Delta=10\%$, 20\%, 40\% are also considered. In the case of small $h^{}_d$, the magnetization decay depends weakly on $\alpha$ and $\Delta$. There is a rapid decay of magnetization for $\alpha>60^\circ$ and significant $h^{}_d$.}
\label{figure6}
\end{figure}

In Fig.~(\ref{figure6}), we study the magnetic relaxation characteristics in a highly anisotropic system $A^{}_r=400.0$. It corresponds to a one-dimensional chain of MNPs. Such a system is of high importance due to its immense application in digital data storage. In the case of small dipolar interaction strength ($h^{}_d=0.2$), the function form of the relaxation curve is exponentially decaying, irrespective of anisotropy axis orientation and disorder strength $\Delta$. Interesting physics unfolds with large dipolar interaction strength. The magnetization ceases to relax for anisotropy orientational angle $\alpha<60^\circ$ and $h^{}_d=0.6$ even with small disorder strength $\Delta$. It is because the shape anisotropy in such a case is the maximum that promotes ferromagnetic coupling among the magnetic moments. The out-of-plane disorder further enhances such ferromagnetic interaction. Interestingly, there is a rapid decay of magnetization for $\alpha>60^\circ$, irrespective of disorder strength $\Delta$. In such cases, the majority of magnetic moments momentarily changes their directions by $180^\circ$, resulting in negative averaged magnetization.
It can be explained by the fact that the dipolar field and anisotropy axis are directed antiparallel to each other. As a result, magnetization prefers to align antiparallel to the anisotropy axis for large dipolar interaction strength. The strong dependence of the magnetic relaxation with $\alpha$ and $h^{}_d$ is in qualitative agreement with the work of 
Laslett {\it et al.} and Hovorka {\it et al.}~\cite{laslett2016,hovorka2014} for perfectly ordered system ($\Delta=0$).
%These results are in perfect aggrement with our recent theoretical and numerical work~\cite{anand2019}.
 
\begin{figure}[!htb]
\centering\includegraphics[scale=0.50]{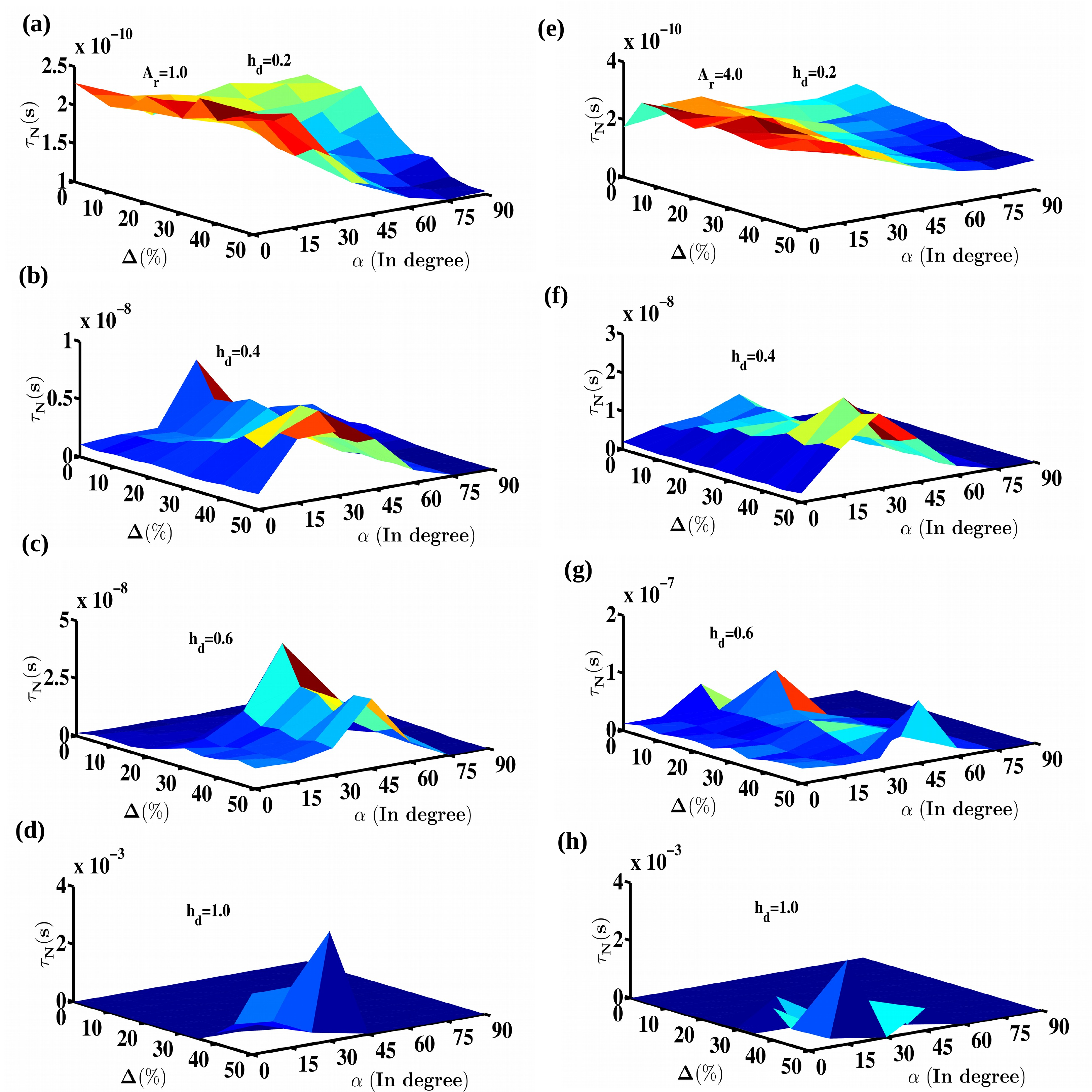}
\caption{The variation of effective N\'eel relaxation time $\tau^{}_N$ as a function of $\alpha$ and $\Delta$. We have considered two values aspect ratio $A^{}_r=1.0$ [(a)-(d)] and 4.0 [(e)-(h)]. We have also considered four values of $h^{}_d=0.2$, 0.4, 0.6, and 1.0. There is a weak dependence of $\tau^{}_N$ on $\alpha$ and $\Delta$ with small $h^{}_d$. $\tau^{}_N$ increases with $\alpha$ ($\leq45^\circ$) for moderate $h^{}_d$ and $A^{}_r=1.0$.}
\label{figure7}
\end{figure}

\begin{figure}[!htb]
	\centering\includegraphics[scale=0.50]{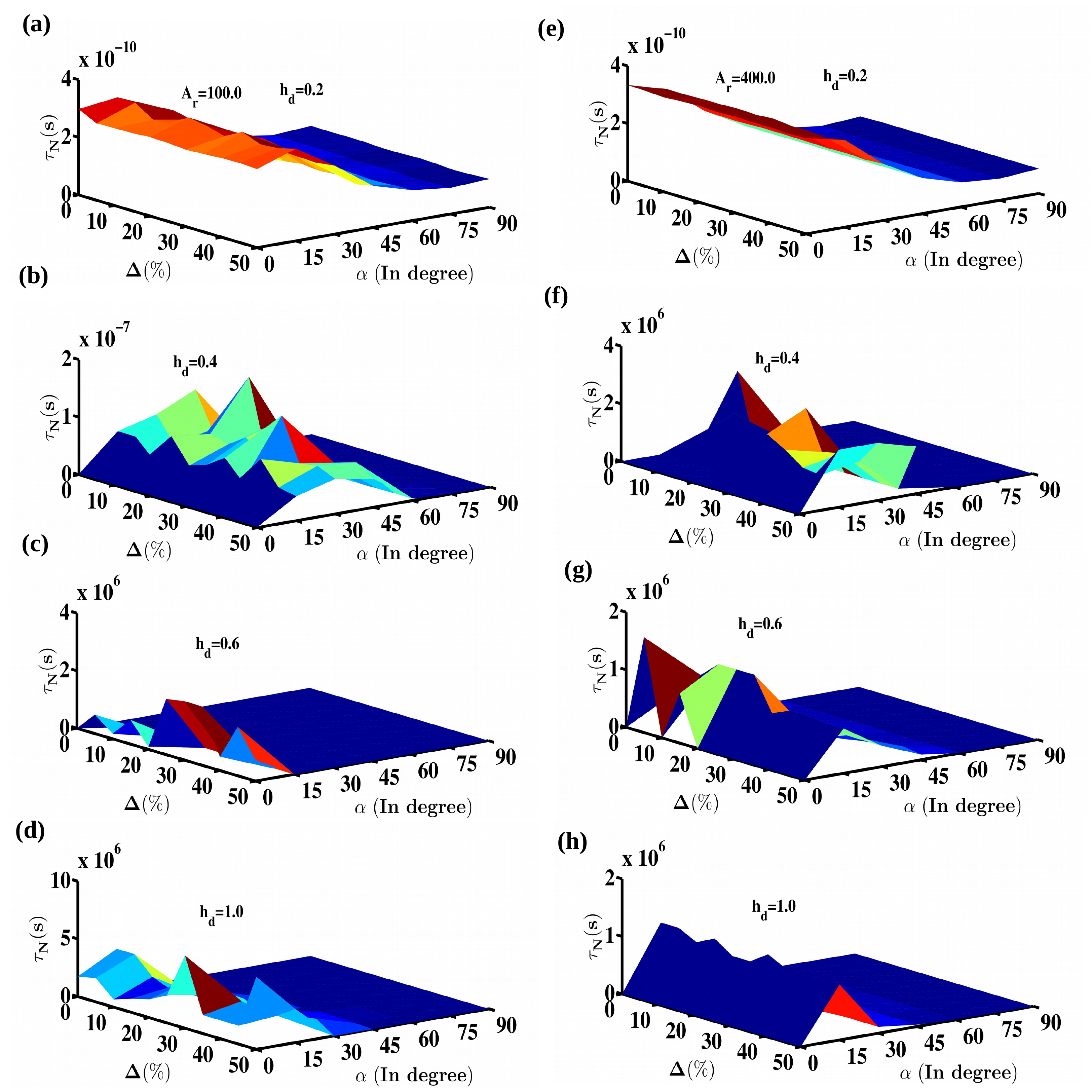}
	\caption{The variation of effective N\'eel relaxation time $\tau^{}_N$ as a function of $\alpha$ and $\Delta$. We have considered two values of aspect ratio $A^{}_r=100.0$ [(a)-(d)], and 400.0 [(e)-(h)]. We have also considered four typical values of $h^{}_d=0.2$, 0.4, 0.6, and 1.0. $\tau^{}_N$ decreases slowly with $\alpha$ for $h^{}_d=0.2$. There is an enhancement in $\tau^{}_N$ with $\alpha$ up to particular $\alpha$ for large $h^{}_d$ and $\Delta$.}
	\label{figure8}
\end{figure}

Finally, we analyze the effective N\'eel relaxation time $\tau^{}_N$ as a function of $\alpha$ and $\Delta$ for various values of dipolar interaction strength $h^{}_d$ and aspect ratio $A^{}_r$ of the system in Fig.~(\ref{figure7}) and Fig.~(\ref{figure8}). In the square-like MNPs assembly, there is a weak dependence of $\tau^{}_N$ on $\alpha$ and $\Delta$ with $h^{}_d=0.2$. Remarkably, $\tau^{}_N$ increases with $\alpha$ ($\leq45^\circ$) for moderate dipolar interaction strength $h^{}_d=0.4$ and 0.6, irrespective of $\Delta$. In contrast, $\tau^{}_N$ decreases with $\alpha$ ($>45^\circ$) in such cases. It is because of antiferromagnetic coupling dominance in square-like MNPs ensemble with the negligible positional disorder. Interestingly, $\tau^{}_N$ increases with $\alpha$ provided $\alpha<45^\circ$ and $\Delta$ is significant. In such cases, the out-of-plane disorder tends to instigate ferromagnetic coupling resulting in the enhancement of $\tau^{}_N$. Remarkably, the antiferromagnetic coupling prevails for smaller $\Delta$ and $\alpha>45^\circ$, which diminishes the value of $\tau^{}_N$. In a highly anisotropic system (huge $A^{}_r$), $\tau^{}_N$ decreases slowly with $\alpha$ even for weakly interacting MNPs $h^{}_d=0.2$. Notably, there is an increase in 
$\tau^{}_N$ increases with $\alpha$ ($\leq30^\circ$) even for moderate interaction strength $h^{}_d$. On the other hand, $\tau^{}_N$ starts to decrease for $\alpha>30^\circ$. The decrease of $\tau^{}_N$ occurs relatively at a smaller value of $\alpha$ ($\leq15^\circ$) for large $h^{}_d$. It is because the dipolar field and anisotropy are antiparallel to each other in such cases, which results in a decrease in relaxation time. The variation of $\tau^{}_N$ with $\alpha$ and $h^{}_d$ for $A_r=400.0$ is in agreement with our earlier work~\cite{anand2019}.  
%\newpage
\section{Summary and Conclusion}
In summary, we have performed kinetic Monte Carlo simulations to investigate magnetic relaxation behaviour in the two-dimensional assembly of MNPs with aligned anisotropy axes and positional defects. Such anisotropic systems find immense utilization in applications such as novel magnetic composite materials, sensors, spintronics based applications, etc.~\cite{song2014,feng2019,castellanos2021}. The variation of the anisotropy axes orientation and disorder strength are controlled by $\alpha$ and $\Delta$, respectively. We have also investigated the relaxation properties as a function of dipolar interaction strength $h^{}_d$ and aspect ratio $A^{}_r$ of the underlying system. The magnetization decays smoothly for weakly interacting MNPs ($h^{}_d=0.2$), irrespective of $A^{}_r$. In such cases, the functional form of the time evolution of the magnetization-decay is also exponentially decaying and  depends weakly on $\alpha$ and $\Delta$. In contrast, there is a strong dependence of relaxation characteristics on the anisotropy axes orientation, out-of-plane disorder strength and aspect ratio for strongly dipolar interacting MNPs. In the square arrays of MNPs and irrespective of $\alpha$, the magnetization decays rapidly for considerable dipolar interaction strength and negligible positional disorder strength $\Delta$. It can be explained by the fact that the dipolar interaction of enough strength promotes antiferromagnetic coupling among the magnetic moments in square ensembles of nanoparticles, resulting in faster magnetization decay. It is in perfect qualitative agreement with the work of MacIsaac {\it et al.}~\cite{macisaac1996}. While with significant disorder strength $\Delta$, the magnetization ceases to relax. Consequently, there is a prolonged magnetization decay for large $\Delta$. The out-of-plane disorder of sufficient strength induces ferromagnetic interaction even in a system with $A^{}_r=1.0$. In a system with a large aspect ratio $A^{}_r$, the magnetization decays extremely slowly for $\alpha<\alpha^\star$ even with moderate interaction strength $h^{}_d$. Interestingly, the slowing down of the magnetic relaxation shifts to a lower value of $\alpha^{\star}$ with the most considerable dipolar interaction strength $h^{}_d=1.0$. Unusual magnetic relaxation behaviour is observed in the highly anisotropic system ($A^{}_r=400.0$) due to interplay among the anisotropy axis orientation, shape anisotropy, and disorder strength. Even in a perfectly ordered system, there is a prolonged magnetization decay for $\alpha\leq60^\circ$ and moderate interaction strength $h^{}_d=0.6$. In such a case, the shape anisotropy alone can induce sufficient ferromagnetic interaction among the magnetic moments. Remarkably, the magnetization decays rapidly for $\alpha>60^\circ$, independent of disorder strength $\Delta$. Most of the magnetic moment reverses its direction by $180^\circ$, resulting in the negative averaged magnetization in these cases. The dipolar field gets aligned antiparallel to the anisotropy axis for large $\alpha$ in such cases~\cite{anand2019}. Consequently, the magnetic moment tends to get aligned along the dipolar field, resulting in negative magnetization. 

The effective N\'eel relaxation time $\tau^{}_N$ also depends strongly on these parameters. There is weak dependence of $\tau^{}_N$ on $\alpha$ and $\Delta$ in a system with small dipolar interaction ($h^{}_d\leq0.2$) irrespective of aspect ratio $A^{}_r$. Notably, there is an enhancement in $\tau^{}_N$ with $\alpha$ ($\leq45^\circ$) for moderate $h^{}_d$ and $A^{}_r=1.0$. On the other hand, $\tau^{}_N$ decreases with $\alpha$ for significant $h^{}_d$ provided $\alpha$ is greater than $45^\circ$. Such behaviour can be explained by the fact that the antiferromagnetic coupling is dominant in square arrays of MNPs in the presence of sufficiently sizeable dipolar interaction strength and negligible positional disorder.  
%Finally, we analyze the effective N\'eel relaxation time $\tau^{}_N$ as a function of $\alpha$ and $\Delta$ for various values of dipolar interaction strength $h^{}_d$ and aspect ratio $A^{}_r$ of the system in Fig.~(\ref{figure7}) and Fig.~(\ref{figure8}). In the square-like MNPs assembly, there is a weak dependence of $\tau^{}_N$ on $\alpha$ and $\Delta$ with $h^{}_d=0.2$. Remarkably, $\tau^{}_N$ increases with $\alpha$ ($\leq45^\circ$) for moderate dipolar interaction strength $h^{}_d=0.4$ and 0.6, irrespective of $\Delta$. In contrast, $\tau^{}_N$ decreases with $\alpha$ ($>45^\circ$) in such cases. It is because of antiferromagnetic coupling dominance in square-like MNPs ensemble with the negligible positional disorder. Interestingly, $\tau^{}_N$ increases with $\alpha$ provided $\alpha<45^\circ$ and $\Delta$ is significant. In such cases, the out-of-plane disorder tends to instigate ferromagnetic coupling resulting in the enhancement of $\tau^{}_N$. Remarkably, the antiferromagnetic coupling prevails for smaller $\Delta$ and $\alpha>45^\circ$, which diminishes the value of $\tau^{}_N$. 
In a highly anisotropic system, there is an increase in $\tau^{}_N$ with $\alpha$ ($\leq30^\circ$) even with moderate interaction strength $h^{}_d$. While for $\alpha>30^\circ$, $\tau^{}_N$ decreases with $\alpha$. The lowering of $\tau^{}_N$ occurs relatively at smaller $\alpha$ with a large $h^{}_d$. Such unusual behaviour is because the dipolar field gets aligned antiparallel to the anisotropy axes for large $\alpha$. Consequently, the magnetization tends to get aligned along the dipolar field for a significantly large $\alpha$, resulting in a decrease in $\tau^{}_N$.
%In a highly anisotropic system (huge $A^{}_r$), $\tau^{}_N$ decreases slowly with $\alpha$ even for weakly interacting MNPs $h^{}_d=0.2$. Notably, there is an increase in  $\tau^{}_N$ increases with $\alpha$ ($\leq30^\circ$) even for moderate interaction strength $h^{}_d$. On the other hand, $\tau^{}_N$ starts to decrease for $\alpha>30^\circ$. The decrease of $\tau^{}_N$ occurs relatively at a smaller value of $\alpha$ ($\leq15^\circ$) for large $h^{}_d$. It is because the dipolar field and anisotropy are antiparallel to each other in such cases, which results in a decrease in relaxation time. 
The observations made in the present work clearly indicate that we can tune the relaxation behaviour and effective N\'eel relaxation time by careful manoeuvring the system parameters such as anisotropy axes, positional disorder, aspect ratio and dipolar interaction strength. These results are beneficial in various applications such as smart materials, sensors, spintronics based applications, etc. 
\section*{DATA AVAILABILITY}
The data that support the findings of this study are available from the corresponding author upon reasonable request.
%\newpage
\bibliography{ref}

\begin{thebibliography}{67}%
\makeatletter
\providecommand \@ifxundefined [1]{%
 \@ifx{#1\undefined}
}%
\providecommand \@ifnum [1]{%
 \ifnum #1\expandafter \@firstoftwo
 \else \expandafter \@secondoftwo
 \fi
}%
\providecommand \@ifx [1]{%
 \ifx #1\expandafter \@firstoftwo
 \else \expandafter \@secondoftwo
 \fi
}%
\providecommand \natexlab [1]{#1}%
\providecommand \enquote  [1]{``#1''}%
\providecommand \bibnamefont  [1]{#1}%
\providecommand \bibfnamefont [1]{#1}%
\providecommand \citenamefont [1]{#1}%
\providecommand \href@noop [0]{\@secondoftwo}%
\providecommand \href [0]{\begingroup \@sanitize@url \@href}%
\providecommand \@href[1]{\@@startlink{#1}\@@href}%
\providecommand \@@href[1]{\endgroup#1\@@endlink}%
\providecommand \@sanitize@url [0]{\catcode `\\12\catcode `\$12\catcode
  `\&12\catcode `\#12\catcode `\^12\catcode `\_12\catcode `\%12\relax}%
\providecommand \@@startlink[1]{}%
\providecommand \@@endlink[0]{}%
\providecommand \url  [0]{\begingroup\@sanitize@url \@url }%
\providecommand \@url [1]{\endgroup\@href {#1}{\urlprefix }}%
\providecommand \urlprefix  [0]{URL }%
\providecommand \Eprint [0]{\href }%
\providecommand \doibase [0]{http://dx.doi.org/}%
\providecommand \selectlanguage [0]{\@gobble}%
\providecommand \bibinfo  [0]{\@secondoftwo}%
\providecommand \bibfield  [0]{\@secondoftwo}%
\providecommand \translation [1]{[#1]}%
\providecommand \BibitemOpen [0]{}%
\providecommand \bibitemStop [0]{}%
\providecommand \bibitemNoStop [0]{.\EOS\space}%
\providecommand \EOS [0]{\spacefactor3000\relax}%
\providecommand \BibitemShut  [1]{\csname bibitem#1\endcsname}%
\let\auto@bib@innerbib\@empty
%</preamble>
\bibitem [{\citenamefont {Leo}\ \emph {et~al.}(2018)\citenamefont {Leo},
  \citenamefont {Holenstein}, \citenamefont {Schildknecht}, \citenamefont
  {Sendetskyi}, \citenamefont {Luetkens}, \citenamefont {Derlet}, \citenamefont
  {Scagnoli}, \citenamefont {Lan{\c{c}}on}, \citenamefont {Mardegan},
  \citenamefont {Prokscha} \emph {et~al.}}]{leo2018}%
  \BibitemOpen
  \bibfield  {author} {\bibinfo {author} {\bibfnamefont {N.}~\bibnamefont
  {Leo}}, \bibinfo {author} {\bibfnamefont {S.}~\bibnamefont {Holenstein}},
  \bibinfo {author} {\bibfnamefont {D.}~\bibnamefont {Schildknecht}}, \bibinfo
  {author} {\bibfnamefont {O.}~\bibnamefont {Sendetskyi}}, \bibinfo {author}
  {\bibfnamefont {H.}~\bibnamefont {Luetkens}}, \bibinfo {author}
  {\bibfnamefont {P.~M.}\ \bibnamefont {Derlet}}, \bibinfo {author}
  {\bibfnamefont {V.}~\bibnamefont {Scagnoli}}, \bibinfo {author}
  {\bibfnamefont {D.}~\bibnamefont {Lan{\c{c}}on}}, \bibinfo {author}
  {\bibfnamefont {J.~R.}\ \bibnamefont {Mardegan}}, \bibinfo {author}
  {\bibfnamefont {T.}~\bibnamefont {Prokscha}},  \emph {et~al.},\ }\href@noop
  {} {\bibfield  {journal} {\bibinfo  {journal} {Nature Communications}\
  }\textbf {\bibinfo {volume} {9}},\ \bibinfo {pages} {1} (\bibinfo {year}
  {2018})}\BibitemShut {NoStop}%
\bibitem [{\citenamefont {Srivastava}\ \emph {et~al.}(2014)\citenamefont
  {Srivastava}, \citenamefont {Nykypanchuk}, \citenamefont {Fukuto},
  \citenamefont {Halverson}, \citenamefont {Tkachenko}, \citenamefont {Yager},\
  and\ \citenamefont {Gang}}]{srivastava2014}%
  \BibitemOpen
  \bibfield  {author} {\bibinfo {author} {\bibfnamefont {S.}~\bibnamefont
  {Srivastava}}, \bibinfo {author} {\bibfnamefont {D.}~\bibnamefont
  {Nykypanchuk}}, \bibinfo {author} {\bibfnamefont {M.}~\bibnamefont {Fukuto}},
  \bibinfo {author} {\bibfnamefont {J.~D.}\ \bibnamefont {Halverson}}, \bibinfo
  {author} {\bibfnamefont {A.~V.}\ \bibnamefont {Tkachenko}}, \bibinfo {author}
  {\bibfnamefont {K.~G.}\ \bibnamefont {Yager}}, \ and\ \bibinfo {author}
  {\bibfnamefont {O.}~\bibnamefont {Gang}},\ }\href@noop {} {\bibfield
  {journal} {\bibinfo  {journal} {Journal of the American Chemical Society}\
  }\textbf {\bibinfo {volume} {136}},\ \bibinfo {pages} {8323} (\bibinfo {year}
  {2014})}\BibitemShut {NoStop}%
\bibitem [{\citenamefont {Gallina}\ and\ \citenamefont
  {Pastor}(2020)}]{gallina2020}%
  \BibitemOpen
  \bibfield  {author} {\bibinfo {author} {\bibfnamefont {D.}~\bibnamefont
  {Gallina}}\ and\ \bibinfo {author} {\bibfnamefont {G.}~\bibnamefont
  {Pastor}},\ }\href@noop {} {\bibfield  {journal} {\bibinfo  {journal}
  {Physical Review X}\ }\textbf {\bibinfo {volume} {10}},\ \bibinfo {pages}
  {021068} (\bibinfo {year} {2020})}\BibitemShut {NoStop}%
\bibitem [{\citenamefont {Puntes}\ \emph {et~al.}(2001)\citenamefont {Puntes},
  \citenamefont {Krishnan},\ and\ \citenamefont {Alivisatos}}]{puntes2001}%
  \BibitemOpen
  \bibfield  {author} {\bibinfo {author} {\bibfnamefont {V.~F.}\ \bibnamefont
  {Puntes}}, \bibinfo {author} {\bibfnamefont {K.~M.}\ \bibnamefont
  {Krishnan}}, \ and\ \bibinfo {author} {\bibfnamefont {P.}~\bibnamefont
  {Alivisatos}},\ }\href@noop {} {\bibfield  {journal} {\bibinfo  {journal}
  {Applied Physics Letters}\ }\textbf {\bibinfo {volume} {78}},\ \bibinfo
  {pages} {2187} (\bibinfo {year} {2001})}\BibitemShut {NoStop}%
\bibitem [{\citenamefont {Moya}\ \emph {et~al.}(2015)\citenamefont {Moya},
  \citenamefont {Iglesias}, \citenamefont {Batlle},\ and\ \citenamefont
  {Labarta}}]{moya2015}%
  \BibitemOpen
  \bibfield  {author} {\bibinfo {author} {\bibfnamefont {C.}~\bibnamefont
  {Moya}}, \bibinfo {author} {\bibfnamefont {O.}~\bibnamefont {Iglesias}},
  \bibinfo {author} {\bibfnamefont {X.}~\bibnamefont {Batlle}}, \ and\ \bibinfo
  {author} {\bibfnamefont {A.}~\bibnamefont {Labarta}},\ }\href@noop {}
  {\bibfield  {journal} {\bibinfo  {journal} {The Journal of Physical Chemistry
  C}\ }\textbf {\bibinfo {volume} {119}},\ \bibinfo {pages} {24142} (\bibinfo
  {year} {2015})}\BibitemShut {NoStop}%
\bibitem [{\citenamefont {Wang}\ and\ \citenamefont
  {Luijten}(2019)}]{wang2019}%
  \BibitemOpen
  \bibfield  {author} {\bibinfo {author} {\bibfnamefont {Z.}~\bibnamefont
  {Wang}}\ and\ \bibinfo {author} {\bibfnamefont {E.}~\bibnamefont {Luijten}},\
  }\href@noop {} {\bibfield  {journal} {\bibinfo  {journal} {Physical Review
  Letters}\ }\textbf {\bibinfo {volume} {123}},\ \bibinfo {pages} {096101}
  (\bibinfo {year} {2019})}\BibitemShut {NoStop}%
\bibitem [{\citenamefont {S{\'a}nchez}\ \emph {et~al.}(2020)\citenamefont
  {S{\'a}nchez}, \citenamefont {Vasilakaki}, \citenamefont {Lee}, \citenamefont
  {Normile}, \citenamefont {Muscas}, \citenamefont {Murgia}, \citenamefont
  {Andersson}, \citenamefont {Singh}, \citenamefont {Mathieu}, \citenamefont
  {Nordblad} \emph {et~al.}}]{sanchez2020}%
  \BibitemOpen
  \bibfield  {author} {\bibinfo {author} {\bibfnamefont {E.~H.}\ \bibnamefont
  {S{\'a}nchez}}, \bibinfo {author} {\bibfnamefont {M.}~\bibnamefont
  {Vasilakaki}}, \bibinfo {author} {\bibfnamefont {S.~S.}\ \bibnamefont {Lee}},
  \bibinfo {author} {\bibfnamefont {P.~S.}\ \bibnamefont {Normile}}, \bibinfo
  {author} {\bibfnamefont {G.}~\bibnamefont {Muscas}}, \bibinfo {author}
  {\bibfnamefont {M.}~\bibnamefont {Murgia}}, \bibinfo {author} {\bibfnamefont
  {M.~S.}\ \bibnamefont {Andersson}}, \bibinfo {author} {\bibfnamefont
  {G.}~\bibnamefont {Singh}}, \bibinfo {author} {\bibfnamefont
  {R.}~\bibnamefont {Mathieu}}, \bibinfo {author} {\bibfnamefont
  {P.}~\bibnamefont {Nordblad}},  \emph {et~al.},\ }\href@noop {} {\bibfield
  {journal} {\bibinfo  {journal} {Chemistry of Materials}\ }\textbf {\bibinfo
  {volume} {32}},\ \bibinfo {pages} {969} (\bibinfo {year} {2020})}\BibitemShut
  {NoStop}%
\bibitem [{\citenamefont {Var{\'o}n}\ \emph {et~al.}(2013)\citenamefont
  {Var{\'o}n}, \citenamefont {Beleggia}, \citenamefont {Kasama}, \citenamefont
  {Harrison}, \citenamefont {Dunin-Borkowski}, \citenamefont {Puntes},\ and\
  \citenamefont {Frandsen}}]{varon2013dipolar}%
  \BibitemOpen
  \bibfield  {author} {\bibinfo {author} {\bibfnamefont {M.}~\bibnamefont
  {Var{\'o}n}}, \bibinfo {author} {\bibfnamefont {M.}~\bibnamefont {Beleggia}},
  \bibinfo {author} {\bibfnamefont {T.}~\bibnamefont {Kasama}}, \bibinfo
  {author} {\bibfnamefont {R.}~\bibnamefont {Harrison}}, \bibinfo {author}
  {\bibfnamefont {R.~E.}\ \bibnamefont {Dunin-Borkowski}}, \bibinfo {author}
  {\bibfnamefont {V.~F.}\ \bibnamefont {Puntes}}, \ and\ \bibinfo {author}
  {\bibfnamefont {C.}~\bibnamefont {Frandsen}},\ }\href@noop {} {\bibfield
  {journal} {\bibinfo  {journal} {Scientific reports}\ }\textbf {\bibinfo
  {volume} {3}},\ \bibinfo {pages} {1} (\bibinfo {year} {2013})}\BibitemShut
  {NoStop}%
\bibitem [{\citenamefont {Peng}\ \emph {et~al.}(2018)\citenamefont {Peng},
  \citenamefont {Fang}, \citenamefont {Li}, \citenamefont {Wang}, \citenamefont
  {Bruck}, \citenamefont {Zhu}, \citenamefont {Zhang}, \citenamefont
  {Takeuchi}, \citenamefont {Marschilok}, \citenamefont {Stach} \emph
  {et~al.}}]{peng2018}%
  \BibitemOpen
  \bibfield  {author} {\bibinfo {author} {\bibfnamefont {L.}~\bibnamefont
  {Peng}}, \bibinfo {author} {\bibfnamefont {Z.}~\bibnamefont {Fang}}, \bibinfo
  {author} {\bibfnamefont {J.}~\bibnamefont {Li}}, \bibinfo {author}
  {\bibfnamefont {L.}~\bibnamefont {Wang}}, \bibinfo {author} {\bibfnamefont
  {A.~M.}\ \bibnamefont {Bruck}}, \bibinfo {author} {\bibfnamefont
  {Y.}~\bibnamefont {Zhu}}, \bibinfo {author} {\bibfnamefont {Y.}~\bibnamefont
  {Zhang}}, \bibinfo {author} {\bibfnamefont {K.~J.}\ \bibnamefont {Takeuchi}},
  \bibinfo {author} {\bibfnamefont {A.~C.}\ \bibnamefont {Marschilok}},
  \bibinfo {author} {\bibfnamefont {E.~A.}\ \bibnamefont {Stach}},  \emph
  {et~al.},\ }\href@noop {} {\bibfield  {journal} {\bibinfo  {journal} {ACS
  nano}\ }\textbf {\bibinfo {volume} {12}},\ \bibinfo {pages} {820} (\bibinfo
  {year} {2018})}\BibitemShut {NoStop}%
\bibitem [{\citenamefont {Anand}(2021{\natexlab{a}})}]{manish2021}%
  \BibitemOpen
  \bibfield  {author} {\bibinfo {author} {\bibfnamefont {M.}~\bibnamefont
  {Anand}},\ }\href {\doibase 10.1007/s12043-021-02222-w} {\bibfield  {journal}
  {\bibinfo  {journal} {Pramana}\ }\textbf {\bibinfo {volume} {95}} (\bibinfo
  {year} {2021}{\natexlab{a}}),\ 10.1007/s12043-021-02222-w}\BibitemShut
  {NoStop}%
\bibitem [{\citenamefont {Farhan}\ \emph {et~al.}(2013)\citenamefont {Farhan},
  \citenamefont {Derlet}, \citenamefont {Kleibert}, \citenamefont {Balan},
  \citenamefont {Chopdekar}, \citenamefont {Wyss}, \citenamefont {Anghinolfi},
  \citenamefont {Nolting},\ and\ \citenamefont {Heyderman}}]{farhan2013}%
  \BibitemOpen
  \bibfield  {author} {\bibinfo {author} {\bibfnamefont {A.}~\bibnamefont
  {Farhan}}, \bibinfo {author} {\bibfnamefont {P.}~\bibnamefont {Derlet}},
  \bibinfo {author} {\bibfnamefont {A.}~\bibnamefont {Kleibert}}, \bibinfo
  {author} {\bibfnamefont {A.}~\bibnamefont {Balan}}, \bibinfo {author}
  {\bibfnamefont {R.}~\bibnamefont {Chopdekar}}, \bibinfo {author}
  {\bibfnamefont {M.}~\bibnamefont {Wyss}}, \bibinfo {author} {\bibfnamefont
  {L.}~\bibnamefont {Anghinolfi}}, \bibinfo {author} {\bibfnamefont
  {F.}~\bibnamefont {Nolting}}, \ and\ \bibinfo {author} {\bibfnamefont
  {L.~J.}\ \bibnamefont {Heyderman}},\ }\href@noop {} {\bibfield  {journal}
  {\bibinfo  {journal} {Nature Physics}\ }\textbf {\bibinfo {volume} {9}},\
  \bibinfo {pages} {375} (\bibinfo {year} {2013})}\BibitemShut {NoStop}%
\bibitem [{\citenamefont {Liu}\ \emph {et~al.}(2002)\citenamefont {Liu},
  \citenamefont {Zhu}, \citenamefont {Hu},\ and\ \citenamefont
  {Liu}}]{liu2002}%
  \BibitemOpen
  \bibfield  {author} {\bibinfo {author} {\bibfnamefont {S.}~\bibnamefont
  {Liu}}, \bibinfo {author} {\bibfnamefont {T.}~\bibnamefont {Zhu}}, \bibinfo
  {author} {\bibfnamefont {R.}~\bibnamefont {Hu}}, \ and\ \bibinfo {author}
  {\bibfnamefont {Z.}~\bibnamefont {Liu}},\ }\href@noop {} {\bibfield
  {journal} {\bibinfo  {journal} {Physical Chemistry Chemical Physics}\
  }\textbf {\bibinfo {volume} {4}},\ \bibinfo {pages} {6059} (\bibinfo {year}
  {2002})}\BibitemShut {NoStop}%
\bibitem [{\citenamefont {Puntes}\ \emph {et~al.}(2004)\citenamefont {Puntes},
  \citenamefont {Gorostiza}, \citenamefont {Aruguete}, \citenamefont {Bastus},\
  and\ \citenamefont {Alivisatos}}]{puntes2004}%
  \BibitemOpen
  \bibfield  {author} {\bibinfo {author} {\bibfnamefont {V.~F.}\ \bibnamefont
  {Puntes}}, \bibinfo {author} {\bibfnamefont {P.}~\bibnamefont {Gorostiza}},
  \bibinfo {author} {\bibfnamefont {D.~M.}\ \bibnamefont {Aruguete}}, \bibinfo
  {author} {\bibfnamefont {N.~G.}\ \bibnamefont {Bastus}}, \ and\ \bibinfo
  {author} {\bibfnamefont {A.~P.}\ \bibnamefont {Alivisatos}},\ }\href@noop {}
  {\bibfield  {journal} {\bibinfo  {journal} {Nature Materials}\ }\textbf
  {\bibinfo {volume} {3}},\ \bibinfo {pages} {263} (\bibinfo {year}
  {2004})}\BibitemShut {NoStop}%
\bibitem [{\citenamefont {Fabris}\ \emph {et~al.}(2019)\citenamefont {Fabris},
  \citenamefont {Lima}, \citenamefont {De~Biasi}, \citenamefont {Troiani},
  \citenamefont {Mansilla}, \citenamefont {Torres}, \citenamefont {Pacheco},
  \citenamefont {Ibarra}, \citenamefont {Goya}, \citenamefont {Zysler} \emph
  {et~al.}}]{fabris2019}%
  \BibitemOpen
  \bibfield  {author} {\bibinfo {author} {\bibfnamefont {F.}~\bibnamefont
  {Fabris}}, \bibinfo {author} {\bibfnamefont {E.}~\bibnamefont {Lima}},
  \bibinfo {author} {\bibfnamefont {E.}~\bibnamefont {De~Biasi}}, \bibinfo
  {author} {\bibfnamefont {H.~E.}\ \bibnamefont {Troiani}}, \bibinfo {author}
  {\bibfnamefont {M.~V.}\ \bibnamefont {Mansilla}}, \bibinfo {author}
  {\bibfnamefont {T.~E.}\ \bibnamefont {Torres}}, \bibinfo {author}
  {\bibfnamefont {R.~F.}\ \bibnamefont {Pacheco}}, \bibinfo {author}
  {\bibfnamefont {M.~R.}\ \bibnamefont {Ibarra}}, \bibinfo {author}
  {\bibfnamefont {G.~F.}\ \bibnamefont {Goya}}, \bibinfo {author}
  {\bibfnamefont {R.~D.}\ \bibnamefont {Zysler}},  \emph {et~al.},\ }\href@noop
  {} {\bibfield  {journal} {\bibinfo  {journal} {Nanoscale}\ }\textbf {\bibinfo
  {volume} {11}},\ \bibinfo {pages} {3164} (\bibinfo {year}
  {2019})}\BibitemShut {NoStop}%
\bibitem [{\citenamefont {Perez}\ \emph {et~al.}(2002)\citenamefont {Perez},
  \citenamefont {Josephson}, \citenamefont {O'Loughlin}, \citenamefont
  {H{\"o}gemann},\ and\ \citenamefont {Weissleder}}]{perez2002}%
  \BibitemOpen
  \bibfield  {author} {\bibinfo {author} {\bibfnamefont {J.~M.}\ \bibnamefont
  {Perez}}, \bibinfo {author} {\bibfnamefont {L.}~\bibnamefont {Josephson}},
  \bibinfo {author} {\bibfnamefont {T.}~\bibnamefont {O'Loughlin}}, \bibinfo
  {author} {\bibfnamefont {D.}~\bibnamefont {H{\"o}gemann}}, \ and\ \bibinfo
  {author} {\bibfnamefont {R.}~\bibnamefont {Weissleder}},\ }\href@noop {}
  {\bibfield  {journal} {\bibinfo  {journal} {Nature biotechnology}\ }\textbf
  {\bibinfo {volume} {20}},\ \bibinfo {pages} {816} (\bibinfo {year}
  {2002})}\BibitemShut {NoStop}%
\bibitem [{\citenamefont {Pisane}\ \emph {et~al.}(2015)\citenamefont {Pisane},
  \citenamefont {Despeaux},\ and\ \citenamefont {Seehra}}]{pisane2015}%
  \BibitemOpen
  \bibfield  {author} {\bibinfo {author} {\bibfnamefont {K.~L.}\ \bibnamefont
  {Pisane}}, \bibinfo {author} {\bibfnamefont {E.~C.}\ \bibnamefont
  {Despeaux}}, \ and\ \bibinfo {author} {\bibfnamefont {M.~S.}\ \bibnamefont
  {Seehra}},\ }\href@noop {} {\bibfield  {journal} {\bibinfo  {journal}
  {Journal of Magnetism and Magnetic Materials}\ }\textbf {\bibinfo {volume}
  {384}},\ \bibinfo {pages} {148} (\bibinfo {year} {2015})}\BibitemShut
  {NoStop}%
\bibitem [{\citenamefont {Hergt}\ \emph {et~al.}(2009)\citenamefont {Hergt},
  \citenamefont {Dutz},\ and\ \citenamefont {Zeisberger}}]{hergt2009}%
  \BibitemOpen
  \bibfield  {author} {\bibinfo {author} {\bibfnamefont {R.}~\bibnamefont
  {Hergt}}, \bibinfo {author} {\bibfnamefont {S.}~\bibnamefont {Dutz}}, \ and\
  \bibinfo {author} {\bibfnamefont {M.}~\bibnamefont {Zeisberger}},\
  }\href@noop {} {\bibfield  {journal} {\bibinfo  {journal} {Nanotechnology}\
  }\textbf {\bibinfo {volume} {21}},\ \bibinfo {pages} {015706} (\bibinfo
  {year} {2009})}\BibitemShut {NoStop}%
\bibitem [{\citenamefont {Wernsdorfer}\ \emph {et~al.}(1997)\citenamefont
  {Wernsdorfer}, \citenamefont {Orozco}, \citenamefont {Hasselbach},
  \citenamefont {Benoit}, \citenamefont {Barbara}, \citenamefont {Demoncy},
  \citenamefont {Loiseau}, \citenamefont {Pascard},\ and\ \citenamefont
  {Mailly}}]{wernsdorfer1997}%
  \BibitemOpen
  \bibfield  {author} {\bibinfo {author} {\bibfnamefont {W.}~\bibnamefont
  {Wernsdorfer}}, \bibinfo {author} {\bibfnamefont {E.~B.}\ \bibnamefont
  {Orozco}}, \bibinfo {author} {\bibfnamefont {K.}~\bibnamefont {Hasselbach}},
  \bibinfo {author} {\bibfnamefont {A.}~\bibnamefont {Benoit}}, \bibinfo
  {author} {\bibfnamefont {B.}~\bibnamefont {Barbara}}, \bibinfo {author}
  {\bibfnamefont {N.}~\bibnamefont {Demoncy}}, \bibinfo {author} {\bibfnamefont
  {A.}~\bibnamefont {Loiseau}}, \bibinfo {author} {\bibfnamefont
  {H.}~\bibnamefont {Pascard}}, \ and\ \bibinfo {author} {\bibfnamefont
  {D.}~\bibnamefont {Mailly}},\ }\href@noop {} {\bibfield  {journal} {\bibinfo
  {journal} {Physical Review Letters}\ }\textbf {\bibinfo {volume} {78}},\
  \bibinfo {pages} {1791} (\bibinfo {year} {1997})}\BibitemShut {NoStop}%
\bibitem [{\citenamefont {Figueiredo}\ and\ \citenamefont
  {Schwarzacher}(2007)}]{figueiredo2007}%
  \BibitemOpen
  \bibfield  {author} {\bibinfo {author} {\bibfnamefont {W.}~\bibnamefont
  {Figueiredo}}\ and\ \bibinfo {author} {\bibfnamefont {W.}~\bibnamefont
  {Schwarzacher}},\ }\href@noop {} {\bibfield  {journal} {\bibinfo  {journal}
  {Journal of Physics: Condensed Matter}\ }\textbf {\bibinfo {volume} {19}},\
  \bibinfo {pages} {276203} (\bibinfo {year} {2007})}\BibitemShut {NoStop}%
\bibitem [{\citenamefont {D{\'e}jardin}(2011)}]{dejardin2011}%
  \BibitemOpen
  \bibfield  {author} {\bibinfo {author} {\bibfnamefont {P.-M.}\ \bibnamefont
  {D{\'e}jardin}},\ }\href@noop {} {\bibfield  {journal} {\bibinfo  {journal}
  {Journal of Applied Physics}\ }\textbf {\bibinfo {volume} {110}},\ \bibinfo
  {pages} {113921} (\bibinfo {year} {2011})}\BibitemShut {NoStop}%
\bibitem [{\citenamefont {Rizzo}\ \emph {et~al.}(1999)\citenamefont {Rizzo},
  \citenamefont {Silva},\ and\ \citenamefont {Kos}}]{rizzo1999}%
  \BibitemOpen
  \bibfield  {author} {\bibinfo {author} {\bibfnamefont {N.}~\bibnamefont
  {Rizzo}}, \bibinfo {author} {\bibfnamefont {T.}~\bibnamefont {Silva}}, \ and\
  \bibinfo {author} {\bibfnamefont {A.}~\bibnamefont {Kos}},\ }\href@noop {}
  {\bibfield  {journal} {\bibinfo  {journal} {Physical Review Letters}\
  }\textbf {\bibinfo {volume} {83}},\ \bibinfo {pages} {4876} (\bibinfo {year}
  {1999})}\BibitemShut {NoStop}%
\bibitem [{\citenamefont {Denisov}\ and\ \citenamefont
  {Trohidou}(2002)}]{denisov2002}%
  \BibitemOpen
  \bibfield  {author} {\bibinfo {author} {\bibfnamefont {S.}~\bibnamefont
  {Denisov}}\ and\ \bibinfo {author} {\bibfnamefont {K.}~\bibnamefont
  {Trohidou}},\ }\href@noop {} {\bibfield  {journal} {\bibinfo  {journal}
  {physica status solidi (a)}\ }\textbf {\bibinfo {volume} {189}},\ \bibinfo
  {pages} {265} (\bibinfo {year} {2002})}\BibitemShut {NoStop}%
\bibitem [{\citenamefont {Hiroi}\ \emph {et~al.}(2011)\citenamefont {Hiroi},
  \citenamefont {Komatsu},\ and\ \citenamefont {Sato}}]{hiroi2011}%
  \BibitemOpen
  \bibfield  {author} {\bibinfo {author} {\bibfnamefont {K.}~\bibnamefont
  {Hiroi}}, \bibinfo {author} {\bibfnamefont {K.}~\bibnamefont {Komatsu}}, \
  and\ \bibinfo {author} {\bibfnamefont {T.}~\bibnamefont {Sato}},\ }\href@noop
  {} {\bibfield  {journal} {\bibinfo  {journal} {Physical Review B}\ }\textbf
  {\bibinfo {volume} {83}},\ \bibinfo {pages} {224423} (\bibinfo {year}
  {2011})}\BibitemShut {NoStop}%
\bibitem [{\citenamefont {Kesserwan}\ \emph {et~al.}(2011)\citenamefont
  {Kesserwan}, \citenamefont {Manfredi}, \citenamefont {Bigot},\ and\
  \citenamefont {Hervieux}}]{kesserwan2011}%
  \BibitemOpen
  \bibfield  {author} {\bibinfo {author} {\bibfnamefont {H.}~\bibnamefont
  {Kesserwan}}, \bibinfo {author} {\bibfnamefont {G.}~\bibnamefont {Manfredi}},
  \bibinfo {author} {\bibfnamefont {J.-Y.}\ \bibnamefont {Bigot}}, \ and\
  \bibinfo {author} {\bibfnamefont {P.-A.}\ \bibnamefont {Hervieux}},\
  }\href@noop {} {\bibfield  {journal} {\bibinfo  {journal} {Physical Review
  B}\ }\textbf {\bibinfo {volume} {84}},\ \bibinfo {pages} {172407} (\bibinfo
  {year} {2011})}\BibitemShut {NoStop}%
\bibitem [{\citenamefont {Kuncser}\ \emph {et~al.}(2019)\citenamefont
  {Kuncser}, \citenamefont {Iacob},\ and\ \citenamefont
  {Kuncser}}]{kuncser2019}%
  \BibitemOpen
  \bibfield  {author} {\bibinfo {author} {\bibfnamefont {A.}~\bibnamefont
  {Kuncser}}, \bibinfo {author} {\bibfnamefont {N.}~\bibnamefont {Iacob}}, \
  and\ \bibinfo {author} {\bibfnamefont {V.~E.}\ \bibnamefont {Kuncser}},\
  }\href@noop {} {\bibfield  {journal} {\bibinfo  {journal} {Beilstein journal
  of nanotechnology}\ }\textbf {\bibinfo {volume} {10}},\ \bibinfo {pages}
  {1280} (\bibinfo {year} {2019})}\BibitemShut {NoStop}%
\bibitem [{\citenamefont {Shtrikman}\ and\ \citenamefont
  {Wohlfarth}(1981)}]{shtrikman1981}%
  \BibitemOpen
  \bibfield  {author} {\bibinfo {author} {\bibfnamefont {S.}~\bibnamefont
  {Shtrikman}}\ and\ \bibinfo {author} {\bibfnamefont {E.}~\bibnamefont
  {Wohlfarth}},\ }\href@noop {} {\bibfield  {journal} {\bibinfo  {journal}
  {Physics Letters A}\ }\textbf {\bibinfo {volume} {85}},\ \bibinfo {pages}
  {467} (\bibinfo {year} {1981})}\BibitemShut {NoStop}%
\bibitem [{\citenamefont {M{\o}rup}\ \emph {et~al.}(2010)\citenamefont
  {M{\o}rup}, \citenamefont {Hansen},\ and\ \citenamefont
  {Frandsen}}]{morup2010}%
  \BibitemOpen
  \bibfield  {author} {\bibinfo {author} {\bibfnamefont {S.}~\bibnamefont
  {M{\o}rup}}, \bibinfo {author} {\bibfnamefont {M.~F.}\ \bibnamefont
  {Hansen}}, \ and\ \bibinfo {author} {\bibfnamefont {C.}~\bibnamefont
  {Frandsen}},\ }\href@noop {} {\bibfield  {journal} {\bibinfo  {journal}
  {Beilstein journal of nanotechnology}\ }\textbf {\bibinfo {volume} {1}},\
  \bibinfo {pages} {182} (\bibinfo {year} {2010})}\BibitemShut {NoStop}%
\bibitem [{\citenamefont {Dormann}\ \emph {et~al.}(1988)\citenamefont
  {Dormann}, \citenamefont {Bessais},\ and\ \citenamefont
  {Fiorani}}]{dormann1988}%
  \BibitemOpen
  \bibfield  {author} {\bibinfo {author} {\bibfnamefont {J.}~\bibnamefont
  {Dormann}}, \bibinfo {author} {\bibfnamefont {L.}~\bibnamefont {Bessais}}, \
  and\ \bibinfo {author} {\bibfnamefont {D.}~\bibnamefont {Fiorani}},\
  }\href@noop {} {\bibfield  {journal} {\bibinfo  {journal} {Journal of Physics
  C: Solid State Physics}\ }\textbf {\bibinfo {volume} {21}},\ \bibinfo {pages}
  {2015} (\bibinfo {year} {1988})}\BibitemShut {NoStop}%
\bibitem [{\citenamefont {Anand}(2021{\natexlab{b}})}]{anand2021thermal}%
  \BibitemOpen
  \bibfield  {author} {\bibinfo {author} {\bibfnamefont {M.}~\bibnamefont
  {Anand}},\ }\href@noop {} {\bibfield  {journal} {\bibinfo  {journal} {Journal
  of Magnetism and Magnetic Materials}\ }\textbf {\bibinfo {volume} {522}},\
  \bibinfo {pages} {167538} (\bibinfo {year} {2021}{\natexlab{b}})}\BibitemShut
  {NoStop}%
\bibitem [{\citenamefont {Salvador}\ \emph {et~al.}(2021)\citenamefont
  {Salvador}, \citenamefont {Nicolao},\ and\ \citenamefont
  {Figueiredo}}]{salvador2021}%
  \BibitemOpen
  \bibfield  {author} {\bibinfo {author} {\bibfnamefont {M.}~\bibnamefont
  {Salvador}}, \bibinfo {author} {\bibfnamefont {L.}~\bibnamefont {Nicolao}}, \
  and\ \bibinfo {author} {\bibfnamefont {W.}~\bibnamefont {Figueiredo}},\
  }\href@noop {} {\bibfield  {journal} {\bibinfo  {journal} {Journal of
  Magnetism and Magnetic Materials}\ }\textbf {\bibinfo {volume} {538}},\
  \bibinfo {pages} {168254} (\bibinfo {year} {2021})}\BibitemShut {NoStop}%
\bibitem [{\citenamefont {Anand}(2021{\natexlab{c}})}]{anand2021magnetic}%
  \BibitemOpen
  \bibfield  {author} {\bibinfo {author} {\bibfnamefont {M.}~\bibnamefont
  {Anand}},\ }\href@noop {} {\bibfield  {journal} {\bibinfo  {journal} {arXiv
  preprint arXiv:2105.00472}\ } (\bibinfo {year}
  {2021}{\natexlab{c}})}\BibitemShut {NoStop}%
\bibitem [{\citenamefont {Ilg}(2017)}]{ilg2017}%
  \BibitemOpen
  \bibfield  {author} {\bibinfo {author} {\bibfnamefont {P.}~\bibnamefont
  {Ilg}},\ }\href@noop {} {\bibfield  {journal} {\bibinfo  {journal} {Physical
  Review B}\ }\textbf {\bibinfo {volume} {95}},\ \bibinfo {pages} {214427}
  (\bibinfo {year} {2017})}\BibitemShut {NoStop}%
\bibitem [{\citenamefont {L{\'o}pez}\ \emph {et~al.}(2001)\citenamefont
  {L{\'o}pez}, \citenamefont {Lisboa-Filho}, \citenamefont {Passos},
  \citenamefont {Ortiz},\ and\ \citenamefont {Araujo-Moreira}}]{lopez2001}%
  \BibitemOpen
  \bibfield  {author} {\bibinfo {author} {\bibfnamefont {J.}~\bibnamefont
  {L{\'o}pez}}, \bibinfo {author} {\bibfnamefont {P.}~\bibnamefont
  {Lisboa-Filho}}, \bibinfo {author} {\bibfnamefont {W.}~\bibnamefont
  {Passos}}, \bibinfo {author} {\bibfnamefont {W.}~\bibnamefont {Ortiz}}, \
  and\ \bibinfo {author} {\bibfnamefont {F.}~\bibnamefont {Araujo-Moreira}},\
  }\href@noop {} {\bibfield  {journal} {\bibinfo  {journal} {Journal of
  magnetism and magnetic materials}\ }\textbf {\bibinfo {volume} {226}},\
  \bibinfo {pages} {500} (\bibinfo {year} {2001})}\BibitemShut {NoStop}%
\bibitem [{\citenamefont {Osaci}\ \emph {et~al.}(2006)\citenamefont {Osaci},
  \citenamefont {P{\u{a}}noiu}, \citenamefont {Hepu{\c{t}}},\ and\
  \citenamefont {Muscalagiu}}]{osaci2006}%
  \BibitemOpen
  \bibfield  {author} {\bibinfo {author} {\bibfnamefont {M.}~\bibnamefont
  {Osaci}}, \bibinfo {author} {\bibfnamefont {M.}~\bibnamefont {P{\u{a}}noiu}},
  \bibinfo {author} {\bibfnamefont {T.}~\bibnamefont {Hepu{\c{t}}}}, \ and\
  \bibinfo {author} {\bibfnamefont {I.}~\bibnamefont {Muscalagiu}},\
  }\href@noop {} {\bibfield  {journal} {\bibinfo  {journal} {Applied
  Mathematical Modelling}\ }\textbf {\bibinfo {volume} {30}},\ \bibinfo {pages}
  {545} (\bibinfo {year} {2006})}\BibitemShut {NoStop}%
\bibitem [{\citenamefont {Eloi}\ \emph {et~al.}(2014)\citenamefont {Eloi},
  \citenamefont {Okuda}, \citenamefont {Carreira}, \citenamefont
  {Schwarzacher}, \citenamefont {Correia},\ and\ \citenamefont
  {Figueiredo}}]{eloi2014}%
  \BibitemOpen
  \bibfield  {author} {\bibinfo {author} {\bibfnamefont {J.-C.}\ \bibnamefont
  {Eloi}}, \bibinfo {author} {\bibfnamefont {M.}~\bibnamefont {Okuda}},
  \bibinfo {author} {\bibfnamefont {S.~C.}\ \bibnamefont {Carreira}}, \bibinfo
  {author} {\bibfnamefont {W.}~\bibnamefont {Schwarzacher}}, \bibinfo {author}
  {\bibfnamefont {M.~J.}\ \bibnamefont {Correia}}, \ and\ \bibinfo {author}
  {\bibfnamefont {W.}~\bibnamefont {Figueiredo}},\ }\href@noop {} {\bibfield
  {journal} {\bibinfo  {journal} {Journal of Physics: Condensed Matter}\
  }\textbf {\bibinfo {volume} {26}},\ \bibinfo {pages} {146006} (\bibinfo
  {year} {2014})}\BibitemShut {NoStop}%
\bibitem [{\citenamefont {Thorkelsson}\ \emph {et~al.}(2015)\citenamefont
  {Thorkelsson}, \citenamefont {Bai},\ and\ \citenamefont
  {Xu}}]{thorkelsson2015}%
  \BibitemOpen
  \bibfield  {author} {\bibinfo {author} {\bibfnamefont {K.}~\bibnamefont
  {Thorkelsson}}, \bibinfo {author} {\bibfnamefont {P.}~\bibnamefont {Bai}}, \
  and\ \bibinfo {author} {\bibfnamefont {T.}~\bibnamefont {Xu}},\ }\href@noop
  {} {\bibfield  {journal} {\bibinfo  {journal} {Nano Today}\ }\textbf
  {\bibinfo {volume} {10}},\ \bibinfo {pages} {48} (\bibinfo {year}
  {2015})}\BibitemShut {NoStop}%
\bibitem [{\citenamefont {Anand}(2020)}]{anand2020hysteresis}%
  \BibitemOpen
  \bibfield  {author} {\bibinfo {author} {\bibfnamefont {M.}~\bibnamefont
  {Anand}},\ }\href@noop {} {\bibfield  {journal} {\bibinfo  {journal} {Journal
  of Applied Physics}\ }\textbf {\bibinfo {volume} {128}},\ \bibinfo {pages}
  {023903} (\bibinfo {year} {2020})}\BibitemShut {NoStop}%
\bibitem [{\citenamefont {Yoshida}\ \emph {et~al.}(2017)\citenamefont
  {Yoshida}, \citenamefont {Matsugi}, \citenamefont {Tsujimura}, \citenamefont
  {Sasayama}, \citenamefont {Enpuku}, \citenamefont {Viereck}, \citenamefont
  {Schilling},\ and\ \citenamefont {Ludwig}}]{yoshida2017effect}%
  \BibitemOpen
  \bibfield  {author} {\bibinfo {author} {\bibfnamefont {T.}~\bibnamefont
  {Yoshida}}, \bibinfo {author} {\bibfnamefont {Y.}~\bibnamefont {Matsugi}},
  \bibinfo {author} {\bibfnamefont {N.}~\bibnamefont {Tsujimura}}, \bibinfo
  {author} {\bibfnamefont {T.}~\bibnamefont {Sasayama}}, \bibinfo {author}
  {\bibfnamefont {K.}~\bibnamefont {Enpuku}}, \bibinfo {author} {\bibfnamefont
  {T.}~\bibnamefont {Viereck}}, \bibinfo {author} {\bibfnamefont
  {M.}~\bibnamefont {Schilling}}, \ and\ \bibinfo {author} {\bibfnamefont
  {F.}~\bibnamefont {Ludwig}},\ }\href@noop {} {\bibfield  {journal} {\bibinfo
  {journal} {Journal of Magnetism and Magnetic Materials}\ }\textbf {\bibinfo
  {volume} {427}},\ \bibinfo {pages} {162} (\bibinfo {year}
  {2017})}\BibitemShut {NoStop}%
\bibitem [{\citenamefont {Wang}(2008)}]{wang2008fept}%
  \BibitemOpen
  \bibfield  {author} {\bibinfo {author} {\bibfnamefont {J.-P.}\ \bibnamefont
  {Wang}},\ }\href@noop {} {\bibfield  {journal} {\bibinfo  {journal}
  {Proceedings of the IEEE}\ }\textbf {\bibinfo {volume} {96}},\ \bibinfo
  {pages} {1847} (\bibinfo {year} {2008})}\BibitemShut {NoStop}%
\bibitem [{\citenamefont {Khurshid}\ \emph {et~al.}(2021)\citenamefont
  {Khurshid}, \citenamefont {Yoosuf}, \citenamefont {Issa}, \citenamefont
  {Attaelmanan},\ and\ \citenamefont {Hadjipanayis}}]{khurshid2021tuning}%
  \BibitemOpen
  \bibfield  {author} {\bibinfo {author} {\bibfnamefont {H.}~\bibnamefont
  {Khurshid}}, \bibinfo {author} {\bibfnamefont {R.}~\bibnamefont {Yoosuf}},
  \bibinfo {author} {\bibfnamefont {B.~A.}\ \bibnamefont {Issa}}, \bibinfo
  {author} {\bibfnamefont {A.~G.}\ \bibnamefont {Attaelmanan}}, \ and\ \bibinfo
  {author} {\bibfnamefont {G.}~\bibnamefont {Hadjipanayis}},\ }\href@noop {}
  {\bibfield  {journal} {\bibinfo  {journal} {Nanomaterials}\ }\textbf
  {\bibinfo {volume} {11}},\ \bibinfo {pages} {3042} (\bibinfo {year}
  {2021})}\BibitemShut {NoStop}%
\bibitem [{\citenamefont {Sharma}\ \emph {et~al.}(2010)\citenamefont {Sharma},
  \citenamefont {Jaffari}, \citenamefont {Shah},\ and\ \citenamefont
  {Pochan}}]{sharma2010}%
  \BibitemOpen
  \bibfield  {author} {\bibinfo {author} {\bibfnamefont {N.}~\bibnamefont
  {Sharma}}, \bibinfo {author} {\bibfnamefont {G.~H.}\ \bibnamefont {Jaffari}},
  \bibinfo {author} {\bibfnamefont {S.~I.}\ \bibnamefont {Shah}}, \ and\
  \bibinfo {author} {\bibfnamefont {D.~J.}\ \bibnamefont {Pochan}},\
  }\href@noop {} {\bibfield  {journal} {\bibinfo  {journal} {Nanotechnology}\
  }\textbf {\bibinfo {volume} {21}},\ \bibinfo {pages} {085707} (\bibinfo
  {year} {2010})}\BibitemShut {NoStop}%
\bibitem [{\citenamefont {Usov}\ \emph {et~al.}(2013)\citenamefont {Usov},
  \citenamefont {Fdez-Gubieda},\ and\ \citenamefont
  {Barandiar{\'a}n}}]{usov2013}%
  \BibitemOpen
  \bibfield  {author} {\bibinfo {author} {\bibfnamefont {N.}~\bibnamefont
  {Usov}}, \bibinfo {author} {\bibfnamefont {M.}~\bibnamefont {Fdez-Gubieda}},
  \ and\ \bibinfo {author} {\bibfnamefont {J.}~\bibnamefont
  {Barandiar{\'a}n}},\ }\href@noop {} {\bibfield  {journal} {\bibinfo
  {journal} {Journal of Applied Physics}\ }\textbf {\bibinfo {volume} {113}},\
  \bibinfo {pages} {023907} (\bibinfo {year} {2013})}\BibitemShut {NoStop}%
\bibitem [{\citenamefont {Conde-Lebor{\'a}n}\ \emph {et~al.}(2015)\citenamefont
  {Conde-Lebor{\'a}n}, \citenamefont {Serantes},\ and\ \citenamefont
  {Baldomir}}]{conde2015}%
  \BibitemOpen
  \bibfield  {author} {\bibinfo {author} {\bibfnamefont {I.}~\bibnamefont
  {Conde-Lebor{\'a}n}}, \bibinfo {author} {\bibfnamefont {D.}~\bibnamefont
  {Serantes}}, \ and\ \bibinfo {author} {\bibfnamefont {D.}~\bibnamefont
  {Baldomir}},\ }\href@noop {} {\bibfield  {journal} {\bibinfo  {journal}
  {Journal of Magnetism and Magnetic Materials}\ }\textbf {\bibinfo {volume}
  {380}},\ \bibinfo {pages} {321} (\bibinfo {year} {2015})}\BibitemShut
  {NoStop}%
\bibitem [{\citenamefont {Anand}\ \emph {et~al.}(2019)\citenamefont {Anand},
  \citenamefont {Banerjee},\ and\ \citenamefont {Carrey}}]{anand2019}%
  \BibitemOpen
  \bibfield  {author} {\bibinfo {author} {\bibfnamefont {M.}~\bibnamefont
  {Anand}}, \bibinfo {author} {\bibfnamefont {V.}~\bibnamefont {Banerjee}}, \
  and\ \bibinfo {author} {\bibfnamefont {J.}~\bibnamefont {Carrey}},\
  }\href@noop {} {\bibfield  {journal} {\bibinfo  {journal} {Physical Review
  B}\ }\textbf {\bibinfo {volume} {99}},\ \bibinfo {pages} {024402} (\bibinfo
  {year} {2019})}\BibitemShut {NoStop}%
\bibitem [{\citenamefont {Anand}(2021{\natexlab{d}})}]{anand2021relaxation}%
  \BibitemOpen
  \bibfield  {author} {\bibinfo {author} {\bibfnamefont {M.}~\bibnamefont
  {Anand}},\ }\href@noop {} {\bibfield  {journal} {\bibinfo  {journal} {arXiv
  preprint arXiv:2106.14271}\ } (\bibinfo {year}
  {2021}{\natexlab{d}})}\BibitemShut {NoStop}%
\bibitem [{\citenamefont {Aldaye}\ \emph {et~al.}(2008)\citenamefont {Aldaye},
  \citenamefont {Palmer},\ and\ \citenamefont {Sleiman}}]{aldaye2008}%
  \BibitemOpen
  \bibfield  {author} {\bibinfo {author} {\bibfnamefont {F.~A.}\ \bibnamefont
  {Aldaye}}, \bibinfo {author} {\bibfnamefont {A.~L.}\ \bibnamefont {Palmer}},
  \ and\ \bibinfo {author} {\bibfnamefont {H.~F.}\ \bibnamefont {Sleiman}},\
  }\href@noop {} {\bibfield  {journal} {\bibinfo  {journal} {science}\ }\textbf
  {\bibinfo {volume} {321}},\ \bibinfo {pages} {1795} (\bibinfo {year}
  {2008})}\BibitemShut {NoStop}%
\bibitem [{\citenamefont {Boal}\ \emph {et~al.}(2000)\citenamefont {Boal},
  \citenamefont {Ilhan}, \citenamefont {DeRouchey}, \citenamefont
  {Thurn-Albrecht}, \citenamefont {Russell},\ and\ \citenamefont
  {Rotello}}]{boal2000}%
  \BibitemOpen
  \bibfield  {author} {\bibinfo {author} {\bibfnamefont {A.~K.}\ \bibnamefont
  {Boal}}, \bibinfo {author} {\bibfnamefont {F.}~\bibnamefont {Ilhan}},
  \bibinfo {author} {\bibfnamefont {J.~E.}\ \bibnamefont {DeRouchey}}, \bibinfo
  {author} {\bibfnamefont {T.}~\bibnamefont {Thurn-Albrecht}}, \bibinfo
  {author} {\bibfnamefont {T.~P.}\ \bibnamefont {Russell}}, \ and\ \bibinfo
  {author} {\bibfnamefont {V.~M.}\ \bibnamefont {Rotello}},\ }\href@noop {}
  {\bibfield  {journal} {\bibinfo  {journal} {Nature}\ }\textbf {\bibinfo
  {volume} {404}},\ \bibinfo {pages} {746} (\bibinfo {year}
  {2000})}\BibitemShut {NoStop}%
\bibitem [{\citenamefont {Huynh}\ \emph {et~al.}(2002)\citenamefont {Huynh},
  \citenamefont {Dittmer},\ and\ \citenamefont {Alivisatos}}]{huynh2002}%
  \BibitemOpen
  \bibfield  {author} {\bibinfo {author} {\bibfnamefont {W.~U.}\ \bibnamefont
  {Huynh}}, \bibinfo {author} {\bibfnamefont {J.~J.}\ \bibnamefont {Dittmer}},
  \ and\ \bibinfo {author} {\bibfnamefont {A.~P.}\ \bibnamefont {Alivisatos}},\
  }\href@noop {} {\bibfield  {journal} {\bibinfo  {journal} {science}\ }\textbf
  {\bibinfo {volume} {295}},\ \bibinfo {pages} {2425} (\bibinfo {year}
  {2002})}\BibitemShut {NoStop}%
\bibitem [{\citenamefont {Fava}\ \emph {et~al.}(2008)\citenamefont {Fava},
  \citenamefont {Nie}, \citenamefont {Winnik},\ and\ \citenamefont
  {Kumacheva}}]{fava2008}%
  \BibitemOpen
  \bibfield  {author} {\bibinfo {author} {\bibfnamefont {D.}~\bibnamefont
  {Fava}}, \bibinfo {author} {\bibfnamefont {Z.}~\bibnamefont {Nie}}, \bibinfo
  {author} {\bibfnamefont {M.~A.}\ \bibnamefont {Winnik}}, \ and\ \bibinfo
  {author} {\bibfnamefont {E.}~\bibnamefont {Kumacheva}},\ }\href@noop {}
  {\bibfield  {journal} {\bibinfo  {journal} {Advanced Materials}\ }\textbf
  {\bibinfo {volume} {20}},\ \bibinfo {pages} {4318} (\bibinfo {year}
  {2008})}\BibitemShut {NoStop}%
\bibitem [{\citenamefont {Ryan}\ \emph {et~al.}(2006)\citenamefont {Ryan},
  \citenamefont {Mastroianni}, \citenamefont {Stancil}, \citenamefont {Liu},\
  and\ \citenamefont {Alivisatos}}]{ryan2006}%
  \BibitemOpen
  \bibfield  {author} {\bibinfo {author} {\bibfnamefont {K.~M.}\ \bibnamefont
  {Ryan}}, \bibinfo {author} {\bibfnamefont {A.}~\bibnamefont {Mastroianni}},
  \bibinfo {author} {\bibfnamefont {K.~A.}\ \bibnamefont {Stancil}}, \bibinfo
  {author} {\bibfnamefont {H.}~\bibnamefont {Liu}}, \ and\ \bibinfo {author}
  {\bibfnamefont {A.}~\bibnamefont {Alivisatos}},\ }\href@noop {} {\bibfield
  {journal} {\bibinfo  {journal} {Nano letters}\ }\textbf {\bibinfo {volume}
  {6}},\ \bibinfo {pages} {1479} (\bibinfo {year} {2006})}\BibitemShut
  {NoStop}%
\bibitem [{\citenamefont {Ploshnik}\ \emph {et~al.}(2010)\citenamefont
  {Ploshnik}, \citenamefont {Salant}, \citenamefont {Banin},\ and\
  \citenamefont {Shenhar}}]{ploshnik2010}%
  \BibitemOpen
  \bibfield  {author} {\bibinfo {author} {\bibfnamefont {E.}~\bibnamefont
  {Ploshnik}}, \bibinfo {author} {\bibfnamefont {A.}~\bibnamefont {Salant}},
  \bibinfo {author} {\bibfnamefont {U.}~\bibnamefont {Banin}}, \ and\ \bibinfo
  {author} {\bibfnamefont {R.}~\bibnamefont {Shenhar}},\ }\href@noop {}
  {\bibfield  {journal} {\bibinfo  {journal} {Advanced Materials}\ }\textbf
  {\bibinfo {volume} {22}},\ \bibinfo {pages} {2774} (\bibinfo {year}
  {2010})}\BibitemShut {NoStop}%
\bibitem [{\citenamefont {Anand}\ \emph {et~al.}(2016)\citenamefont {Anand},
  \citenamefont {Carrey},\ and\ \citenamefont {Banerjee}}]{anand2016spin}%
  \BibitemOpen
  \bibfield  {author} {\bibinfo {author} {\bibfnamefont {M.}~\bibnamefont
  {Anand}}, \bibinfo {author} {\bibfnamefont {J.}~\bibnamefont {Carrey}}, \
  and\ \bibinfo {author} {\bibfnamefont {V.}~\bibnamefont {Banerjee}},\
  }\href@noop {} {\bibfield  {journal} {\bibinfo  {journal} {Physical Review
  B}\ }\textbf {\bibinfo {volume} {94}},\ \bibinfo {pages} {094425} (\bibinfo
  {year} {2016})}\BibitemShut {NoStop}%
\bibitem [{\citenamefont {Carrey}\ \emph {et~al.}(2011)\citenamefont {Carrey},
  \citenamefont {Mehdaoui},\ and\ \citenamefont {Respaud}}]{carrey2011}%
  \BibitemOpen
  \bibfield  {author} {\bibinfo {author} {\bibfnamefont {J.}~\bibnamefont
  {Carrey}}, \bibinfo {author} {\bibfnamefont {B.}~\bibnamefont {Mehdaoui}}, \
  and\ \bibinfo {author} {\bibfnamefont {M.}~\bibnamefont {Respaud}},\
  }\href@noop {} {\bibfield  {journal} {\bibinfo  {journal} {Journal of Applied
  Physics}\ }\textbf {\bibinfo {volume} {109}},\ \bibinfo {pages} {083921}
  (\bibinfo {year} {2011})}\BibitemShut {NoStop}%
\bibitem [{\citenamefont {Anand}(2021{\natexlab{e}})}]{ANAND2021168461}%
  \BibitemOpen
  \bibfield  {author} {\bibinfo {author} {\bibfnamefont {M.}~\bibnamefont
  {Anand}},\ }\href {\doibase https://doi.org/10.1016/j.jmmm.2021.168461}
  {\bibfield  {journal} {\bibinfo  {journal} {Journal of Magnetism and Magnetic
  Materials}\ }\textbf {\bibinfo {volume} {540}},\ \bibinfo {pages} {168461}
  (\bibinfo {year} {2021}{\natexlab{e}})}\BibitemShut {NoStop}%
\bibitem [{\citenamefont {Bupathy}\ \emph {et~al.}(2019)\citenamefont
  {Bupathy}, \citenamefont {Banerjee},\ and\ \citenamefont
  {Carrey}}]{bupathy2019}%
  \BibitemOpen
  \bibfield  {author} {\bibinfo {author} {\bibfnamefont {A.}~\bibnamefont
  {Bupathy}}, \bibinfo {author} {\bibfnamefont {V.}~\bibnamefont {Banerjee}}, \
  and\ \bibinfo {author} {\bibfnamefont {J.}~\bibnamefont {Carrey}},\
  }\href@noop {} {\bibfield  {journal} {\bibinfo  {journal} {Physical Review
  B}\ }\textbf {\bibinfo {volume} {100}},\ \bibinfo {pages} {064420} (\bibinfo
  {year} {2019})}\BibitemShut {NoStop}%
\bibitem [{\citenamefont {Tan}\ \emph {et~al.}(2014)\citenamefont {Tan},
  \citenamefont {Carrey},\ and\ \citenamefont {Respaud}}]{tan2014}%
  \BibitemOpen
  \bibfield  {author} {\bibinfo {author} {\bibfnamefont {R.}~\bibnamefont
  {Tan}}, \bibinfo {author} {\bibfnamefont {J.}~\bibnamefont {Carrey}}, \ and\
  \bibinfo {author} {\bibfnamefont {M.}~\bibnamefont {Respaud}},\ }\href@noop
  {} {\bibfield  {journal} {\bibinfo  {journal} {Physical Review B}\ }\textbf
  {\bibinfo {volume} {90}},\ \bibinfo {pages} {214421} (\bibinfo {year}
  {2014})}\BibitemShut {NoStop}%
\bibitem [{\citenamefont {Azeggagh}\ and\ \citenamefont
  {Kachkachi}(2007)}]{azeggagh2007}%
  \BibitemOpen
  \bibfield  {author} {\bibinfo {author} {\bibfnamefont {M.}~\bibnamefont
  {Azeggagh}}\ and\ \bibinfo {author} {\bibfnamefont {H.}~\bibnamefont
  {Kachkachi}},\ }\href@noop {} {\bibfield  {journal} {\bibinfo  {journal}
  {Physical Review B}\ }\textbf {\bibinfo {volume} {75}},\ \bibinfo {pages}
  {174410} (\bibinfo {year} {2007})}\BibitemShut {NoStop}%
\bibitem [{\citenamefont {Tan}\ \emph {et~al.}(2010)\citenamefont {Tan},
  \citenamefont {Lee}, \citenamefont {Cho}, \citenamefont {Noh}, \citenamefont
  {Kim},\ and\ \citenamefont {Kim}}]{tan2010}%
  \BibitemOpen
  \bibfield  {author} {\bibinfo {author} {\bibfnamefont {R.}~\bibnamefont
  {Tan}}, \bibinfo {author} {\bibfnamefont {J.}~\bibnamefont {Lee}}, \bibinfo
  {author} {\bibfnamefont {J.}~\bibnamefont {Cho}}, \bibinfo {author}
  {\bibfnamefont {S.}~\bibnamefont {Noh}}, \bibinfo {author} {\bibfnamefont
  {D.}~\bibnamefont {Kim}}, \ and\ \bibinfo {author} {\bibfnamefont
  {Y.}~\bibnamefont {Kim}},\ }\href@noop {} {\bibfield  {journal} {\bibinfo
  {journal} {Journal of Physics D: Applied Physics}\ }\textbf {\bibinfo
  {volume} {43}},\ \bibinfo {pages} {165002} (\bibinfo {year}
  {2010})}\BibitemShut {NoStop}%
\bibitem [{\citenamefont {H{\"a}nggi}\ \emph {et~al.}(1990)\citenamefont
  {H{\"a}nggi}, \citenamefont {Talkner},\ and\ \citenamefont
  {Borkovec}}]{hanggi1990}%
  \BibitemOpen
  \bibfield  {author} {\bibinfo {author} {\bibfnamefont {P.}~\bibnamefont
  {H{\"a}nggi}}, \bibinfo {author} {\bibfnamefont {P.}~\bibnamefont {Talkner}},
  \ and\ \bibinfo {author} {\bibfnamefont {M.}~\bibnamefont {Borkovec}},\
  }\href@noop {} {\bibfield  {journal} {\bibinfo  {journal} {Reviews of Modern
  Physics}\ }\textbf {\bibinfo {volume} {62}},\ \bibinfo {pages} {251}
  (\bibinfo {year} {1990})}\BibitemShut {NoStop}%
\bibitem [{\citenamefont {Anand}(2021{\natexlab{f}})}]{anand2021tailoring}%
  \BibitemOpen
  \bibfield  {author} {\bibinfo {author} {\bibfnamefont {M.}~\bibnamefont
  {Anand}},\ }\href@noop {} {\bibfield  {journal} {\bibinfo  {journal} {Nano}\
  }\textbf {\bibinfo {volume} {16}},\ \bibinfo {pages} {2150104} (\bibinfo
  {year} {2021}{\natexlab{f}})}\BibitemShut {NoStop}%
\bibitem [{\citenamefont {De'Bell}\ \emph {et~al.}(1997)\citenamefont
  {De'Bell}, \citenamefont {MacIsaac}, \citenamefont {Booth},\ and\
  \citenamefont {Whitehead}}]{de1997}%
  \BibitemOpen
  \bibfield  {author} {\bibinfo {author} {\bibfnamefont {K.}~\bibnamefont
  {De'Bell}}, \bibinfo {author} {\bibfnamefont {A.}~\bibnamefont {MacIsaac}},
  \bibinfo {author} {\bibfnamefont {I.}~\bibnamefont {Booth}}, \ and\ \bibinfo
  {author} {\bibfnamefont {J.}~\bibnamefont {Whitehead}},\ }\href@noop {}
  {\bibfield  {journal} {\bibinfo  {journal} {Physical Review B}\ }\textbf
  {\bibinfo {volume} {55}},\ \bibinfo {pages} {15108} (\bibinfo {year}
  {1997})}\BibitemShut {NoStop}%
\bibitem [{\citenamefont {Laslett}\ \emph {et~al.}(2016)\citenamefont
  {Laslett}, \citenamefont {Ruta}, \citenamefont {Chantrell}, \citenamefont
  {Barker}, \citenamefont {Friedman},\ and\ \citenamefont
  {Hovorka}}]{laslett2016}%
  \BibitemOpen
  \bibfield  {author} {\bibinfo {author} {\bibfnamefont {O.}~\bibnamefont
  {Laslett}}, \bibinfo {author} {\bibfnamefont {S.}~\bibnamefont {Ruta}},
  \bibinfo {author} {\bibfnamefont {R.}~\bibnamefont {Chantrell}}, \bibinfo
  {author} {\bibfnamefont {J.}~\bibnamefont {Barker}}, \bibinfo {author}
  {\bibfnamefont {G.}~\bibnamefont {Friedman}}, \ and\ \bibinfo {author}
  {\bibfnamefont {O.}~\bibnamefont {Hovorka}},\ }\href@noop {} {\bibfield
  {journal} {\bibinfo  {journal} {Physica B: Condensed Matter}\ }\textbf
  {\bibinfo {volume} {486}},\ \bibinfo {pages} {173} (\bibinfo {year}
  {2016})}\BibitemShut {NoStop}%
\bibitem [{\citenamefont {Hovorka}\ \emph {et~al.}(2014)\citenamefont
  {Hovorka}, \citenamefont {Barker}, \citenamefont {Friedman},\ and\
  \citenamefont {Chantrell}}]{hovorka2014}%
  \BibitemOpen
  \bibfield  {author} {\bibinfo {author} {\bibfnamefont {O.}~\bibnamefont
  {Hovorka}}, \bibinfo {author} {\bibfnamefont {J.}~\bibnamefont {Barker}},
  \bibinfo {author} {\bibfnamefont {G.}~\bibnamefont {Friedman}}, \ and\
  \bibinfo {author} {\bibfnamefont {R.}~\bibnamefont {Chantrell}},\ }\href@noop
  {} {\bibfield  {journal} {\bibinfo  {journal} {Physical Review B}\ }\textbf
  {\bibinfo {volume} {89}},\ \bibinfo {pages} {104410} (\bibinfo {year}
  {2014})}\BibitemShut {NoStop}%
\bibitem [{\citenamefont {Song}\ \emph {et~al.}(2014)\citenamefont {Song},
  \citenamefont {Spencer}, \citenamefont {Jander}, \citenamefont {Nielsen},
  \citenamefont {Stasiak}, \citenamefont {Kasperchik},\ and\ \citenamefont
  {Dhagat}}]{song2014}%
  \BibitemOpen
  \bibfield  {author} {\bibinfo {author} {\bibfnamefont {H.}~\bibnamefont
  {Song}}, \bibinfo {author} {\bibfnamefont {J.}~\bibnamefont {Spencer}},
  \bibinfo {author} {\bibfnamefont {A.}~\bibnamefont {Jander}}, \bibinfo
  {author} {\bibfnamefont {J.}~\bibnamefont {Nielsen}}, \bibinfo {author}
  {\bibfnamefont {J.}~\bibnamefont {Stasiak}}, \bibinfo {author} {\bibfnamefont
  {V.}~\bibnamefont {Kasperchik}}, \ and\ \bibinfo {author} {\bibfnamefont
  {P.}~\bibnamefont {Dhagat}},\ }\href@noop {} {\bibfield  {journal} {\bibinfo
  {journal} {Journal of Applied Physics}\ }\textbf {\bibinfo {volume} {115}},\
  \bibinfo {pages} {17E308} (\bibinfo {year} {2014})}\BibitemShut {NoStop}%
\bibitem [{\citenamefont {Feng}\ \emph {et~al.}(2019)\citenamefont {Feng},
  \citenamefont {Yang}, \citenamefont {Wang}, \citenamefont {Lyu},
  \citenamefont {Li},\ and\ \citenamefont {Yin}}]{feng2019}%
  \BibitemOpen
  \bibfield  {author} {\bibinfo {author} {\bibfnamefont {J.}~\bibnamefont
  {Feng}}, \bibinfo {author} {\bibfnamefont {F.}~\bibnamefont {Yang}}, \bibinfo
  {author} {\bibfnamefont {X.}~\bibnamefont {Wang}}, \bibinfo {author}
  {\bibfnamefont {F.}~\bibnamefont {Lyu}}, \bibinfo {author} {\bibfnamefont
  {Z.}~\bibnamefont {Li}}, \ and\ \bibinfo {author} {\bibfnamefont
  {Y.}~\bibnamefont {Yin}},\ }\href@noop {} {\bibfield  {journal} {\bibinfo
  {journal} {Advanced Materials}\ }\textbf {\bibinfo {volume} {31}},\ \bibinfo
  {pages} {1900789} (\bibinfo {year} {2019})}\BibitemShut {NoStop}%
\bibitem [{\citenamefont {Castellanos}\ \emph {et~al.}(2021)\citenamefont
  {Castellanos}, \citenamefont {Bharti},\ and\ \citenamefont
  {Velev}}]{castellanos2021}%
  \BibitemOpen
  \bibfield  {author} {\bibinfo {author} {\bibfnamefont {N.~I.}\ \bibnamefont
  {Castellanos}}, \bibinfo {author} {\bibfnamefont {B.}~\bibnamefont {Bharti}},
  \ and\ \bibinfo {author} {\bibfnamefont {O.~D.}\ \bibnamefont {Velev}},\
  }\href@noop {} {\bibfield  {journal} {\bibinfo  {journal} {The Journal of
  Physical Chemistry B}\ }\textbf {\bibinfo {volume} {125}},\ \bibinfo {pages}
  {7900} (\bibinfo {year} {2021})}\BibitemShut {NoStop}%
\bibitem [{\citenamefont {MacIsaac}\ \emph {et~al.}(1996)\citenamefont
  {MacIsaac}, \citenamefont {Whitehead}, \citenamefont {De'Bell},\ and\
  \citenamefont {Poole}}]{macisaac1996}%
  \BibitemOpen
  \bibfield  {author} {\bibinfo {author} {\bibfnamefont {A.}~\bibnamefont
  {MacIsaac}}, \bibinfo {author} {\bibfnamefont {J.}~\bibnamefont {Whitehead}},
  \bibinfo {author} {\bibfnamefont {K.}~\bibnamefont {De'Bell}}, \ and\
  \bibinfo {author} {\bibfnamefont {P.}~\bibnamefont {Poole}},\ }\href@noop {}
  {\bibfield  {journal} {\bibinfo  {journal} {Physical review letters}\
  }\textbf {\bibinfo {volume} {77}},\ \bibinfo {pages} {739} (\bibinfo {year}
  {1996})}\BibitemShut {NoStop}%
\end{thebibliography}%
\newpage
\newpage
\newpage

\newpage

\newpage

\newpage
\newpage
%\end{document}
\newpage
\end{document}